\documentclass[twocolumn,showpacs,preprintnumbers,amsmath,amssymb]{revtex4}
\usepackage[dvips]{graphicx} 
\begin{document}

\preprint{IFF-RCA-05-11}
\title{Quantum entanglement in the multiverse}
\author{Salvador Robles-P\'{e}rez and Pedro F. Gonz\'alez-D\'{\i}az}
\affiliation{Colina de los Chopos, Centro de F\'{\i}sica ``Miguel Catal\'{a}n'',
Instituto de F\'{\i}sica Fundamental,
Consejo Superior de Investigaciones Cient\'{\i}ficas, Serrano 121, 28006
Madrid (SPAIN) and \\ Estaci\'{o}n Ecol\'{o}gica de Biocosmolog\'{\i}a, Pedro de Alvarado, 14, 06411-Medell\'{\i}n, (SPAIN).}
\date{\today}

\begin{abstract}
In this paper it is shown that the quantum state of a multiverse made up of classically disconnected regions of the space-time, whose dynamical evolution is dominated by a homogeneous and isotropic fluid, is given by a squeezed state. These are typical quantum states that have no classical counterpart and, therefore, they allow us to analyze the violation of classical inequalities as well as the EPR argument in the context of the quantum multiverse. The thermodynamical properties of entanglement are calculated for a composite quantum state of two universes whose  states are quantum mechanically correlated. The energy of entanglement between the positive and negative modes of a scalar field, which correspond to the expanding and contracting branches of a phantom universe, respectively, are also computed.
\end{abstract}

\maketitle

\section{Introduction}

In quantum optics, there are quantum states that violate some inequalities which should be satisfied in the classical theory of light. The effect of  photon antibunching, the violation of the Cauchy-Schwartz inequality and mainly the violation of Bell's inequalities, clearly reveal   the corpuscular nature of the photon and  the existence of non-local correlations in the quantum state of an entangled pair of photons \cite{Reid1986}. Then, quantum states with no classical analog allow us to understand the concept of complementarity and  the non-local character of the quantum theory.

In the context of a quantum multiverse, the existence of such states  and the violation of classical equations might also help us to understand cosmical effects that have no classical counterpart, one of these could well be the current accelerated expansion of the universe. Furthermore, in the quantum multiverse, which is the natural context of the quantum description of the universe, there is no common space-time among the universes and, therefore,  the concepts of complementarity and non-locality have to be revised or extended. One of the aims of this paper is the analysis of such an extension.

The other aim  is  to study the thermodynamics of a pair of universes whose quantum mechanical states are entangled. As it is well known, entanglement is a quantum feature without classical analog. Actually, gravitational and cosmic entanglement are clearly related to quantum effects that have no classical counterpart as they can be related to the origin of  the black hole thermodynamics \cite{Mukohyama1997, Mukohyama1998}, and, on cosmological grounds, to the current accelerated expansion of the universe \cite{Lee2007,Muller1995} and in the context of a multiverse scenario \cite{Mersini2008a, Mersini2008b}. Therefore, entanglement between the states of two different universes will presumably provide us also with quantum effects having no classical analog.

Quantum entanglement between the states of two universes has to be considered in the multiverse provided that the concept of complementarity can also be applied in  quantum cosmology. In that case, interference processes between the  states of the universes should be  expected, and they might have dynamical and thermodynamical consequences in each single universe.

On the other hand, the context of a quantum multiverse impels us to the introduction of new statistical boundary conditions. These can be given, for instance, by imposing a constant number of universes in the multiverse, a constant energy or a constant entropy, conditions that can be partially determined by the choice of representation which is taken to describe the state of single universes. They can also determine the degree of entanglement between different universes.

The outline of the paper is the following. In Sec. II, we calculate the quantum state of the universe and we conclude that entangled states, squeezed states and generally other quantum states with no classical analog have to be considered in the quantum multiverse. The possible violation of classical inequalities as well as the analog of the EPR argument in the multiverse are also analyzed. In Sec. III, we define the thermodynamical properties of a closed system in the context of a quantum theory of information. Then, we consider a pair of universes whose quantum states are correlated and we calculate the thermodynamics of entanglement in each single universe of the pair. In Sec. IV, we draw some conclusions and further comments.

\section{Quantum states in the multiverse with no classical analog}

\subsection{Third quantization}

The natural formulation of the quantum multiverse is a third quantization scheme \cite{Strominger1990}, where creation and annihilation operators of universes can be defined \cite{Strominger1990, RP2010}. The basic idea is to consider the wave function of the universe as a field that propagates in the superspace of geometries and matter fields and, then, to study the state of the multiverse as a quantum field theory in the superspace. Such a quantum field theory is not well-defined in a general superspace. However, in the case of a homogeneous and isotropic space-time, the Wheeler-DeWitt equation can be interpreted as a Klein-Gordon equation in the minisuperspace defined by the variables $(a,\vec{\varphi})$, where the scale factor, $a$, formally plays the role of the time variable and the scalar fields, $\vec{\varphi}\equiv (\varphi_1, \ldots, \varphi_n)$, play the role of the spatial coordinates in the analogy between the third quantization formalism and the quantum field theory in a curved space-time. The role of the scale factor as the time variable within a single universe can generally be a problematic task (for the customary discussions on the subject, see Refs. \cite{Vilenkin1986, Isham1992, Hawking1992, Hartle1993, Kiefer1995, Kiefer2005, Kiefer2007}). Howerver, in the third quantization formalism, the scale factor is just formally taken as an intrinsic time variable which is allowed by the Lorentzian signature of the minisupermetric of the minisuperspace being considered \cite{DeWitt1967, Strominger1990, Wiltshire2003, Kiefer2007}.

Let us therefore start with  a Friedmann-Robertson-Walker space-time, with closed spatial sections,  which is filled with a homogeneous and isotropic fluid with equation of state given by, $p = w \rho$, where $p$ and $\rho$ are the pressure and the energy density of the fluid, respectively, and $w$ a constant parameter. Let us also consider a massless scalar field, $\varphi$. Then, with an appropriate factor ordering choice \cite{Kiefer2007} and rescaling the scalar field to absorb unimportant constants, the Wheeler-DeWitt equation can be written as
\begin{equation}\label{eq01}
\left( a^2 \partial^2_{aa} + a \partial_a + \frac{\omega_0^2 a^{2 q} - a^4}{\hbar^2}  - \partial^2_{\varphi \varphi} \right) \phi(a, \varphi) = 0 ,
\end{equation}
where $\phi(a, \varphi)$ is the wave function of the universe, which is defined in the minisuperspace spanned by the variables $(a,\varphi)$, and $\omega_0$ is a constant of integration that depends on the energy density of the universe at a given spatial hypersurface, $\Sigma(a_0)$. In Eq. (\ref{eq01}), $q\equiv \frac{3}{2}(1-w)$, just  effectively parameterizes   different kind of fluids that can permeate the universe. For instance, the values $w=\frac{1}{3}$, $w= 0$, and $w= -1$, are  equivalent to consider that the classical evolution of the universe is dominated by a radiation-like fluid, a matter (dust), or a vacuum energy, respectively. For concreteness, let us assume the value $w=-1$  which corresponds to a universe endorsed with a cosmological constant, i.e. $q=3$ and $\omega_0^2 \equiv \Lambda$ in Eq. (\ref{eq01}). Then, the Wheeler-DeWitt equation can be written as
\begin{equation}\label{eq01b}
\hbar^2 \ddot{\phi} + \frac{\hbar^2}{a}  \dot{\phi} -\frac{\hbar^2}{a^2} \phi'' + (\Lambda a^4 - a^2) \phi = 0 ,
\end{equation}
where, $\phi\equiv \phi(a,\varphi)$, $\dot{\phi}\equiv\frac{\partial \phi}{\partial a}$, and $\phi'\equiv\frac{\partial \phi}{\partial \varphi}$. 

In the third quantization formalism, the wave function of the universe is promoted to an operator that can be decomposed in normal modes as
\begin{equation}\label{decomposition}
\hat{\phi}(a,\varphi) = \int d k e^{i k \varphi} A_k(a) \hat{c}_k^\dag + e^{-i k\varphi} A^*_k(a) \hat{c}_k ,
\end{equation}
where the probability amplitude $A_k(a)$ satisfies the equation of the damped harmonic oscillator,
\begin{equation}\label{equationModes}
\hbar^2 \ddot{A}_k(a) + \frac{\hbar^2}{a} \dot{A}_k(a) + \omega_k^2(a) A_k(a) = 0 ,
\end{equation}
with a frequency given by
\begin{equation}\label{Frequency0}
\omega_k(a) = \sqrt{\Lambda a^4 - a^2 + \frac{\hbar^2 k^2}{a^2}} .
\end{equation}
The operators, $\hat{c}\equiv \sqrt{\frac{\omega_{0k}}{2\hbar}} (\hat{\phi} + \frac{i}{\omega_{0k}} \hat{p}_\phi)$ and $\hat{c}^\dag \equiv \sqrt{\frac{\omega_{0k}}{2\hbar}} (\hat{\phi} - \frac{i}{\omega_{0k}} \hat{p}_\phi)$, in Eq. (\ref{decomposition}) are, respectively, the annihilation and creation operators of modes of the universe with $\omega_{0k}$ given by Eq. (\ref{Frequency0}) evaluated at the boundary hypersurface $\Sigma(a_0)$.

\begin{figure}
\includegraphics[width=5.3cm,height=4.4cm]{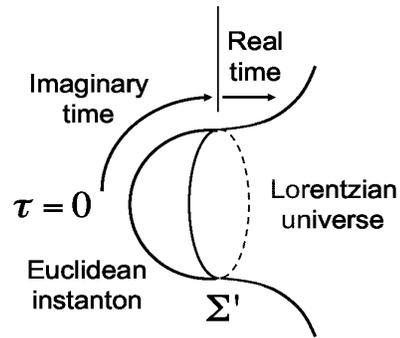}                

\caption{The creation of a De-Sitter universe from a De-Sitter instanton.}
\label{DS_instanton}
\end{figure}

The real values of the frequency (\ref{Frequency0}) define the oscillatory regime of the wave function of the universe in the Lorentzian region, and the complex values define the Euclidean region. Let us first consider the zero mode wave function, i.e. $k=0$. Then, for the value $a > a_+ \equiv \frac{1}{\sqrt{\Lambda}}$, the solution of the Friedmann equation describes the evolution of a closed DeSitter space-time with an eventual exponential expansion of the scale factor with respect to the Friedmann time, i.e. $a(t) \sim e^{\sqrt{\Lambda}t}$. For the value, $a < a_+$, the solution of the Euclidean Friedmann equation corresponds to a DeSitter instanton that eventually collapses at the Euclidean time $\tau = 0$ (see, Fig. \ref{DS_instanton}). This is the customary picture of a DeSitter universe created from a DeSitter instanton \cite{Vilenkin1982, Hawking1983, Vilenkin1986, Kiefer2007}.

\begin{figure}
\includegraphics[width=4.8cm,height=3.8cm]{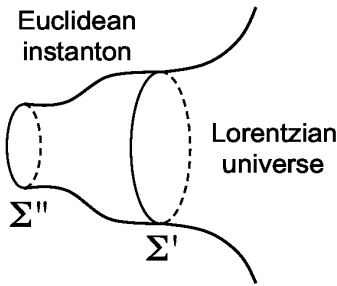}                

\caption{Before reaching the collapse, the instanton finds the transition hypersurface $\Sigma''$.}
\label{EE_instanton}
\end{figure}

\begin{figure}
\includegraphics[width=8cm,height=4cm]{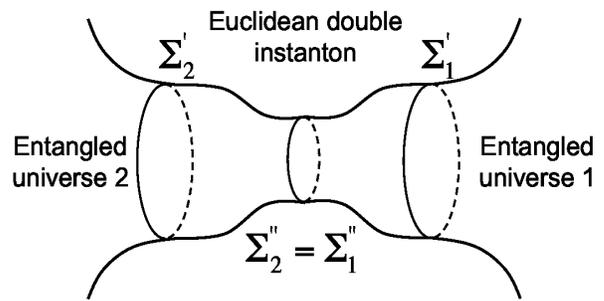}                

\caption{The creation of a pair of entangled universes from a pair of instantons.}
\label{EntangledInstanton}
\end{figure}

The quantum correction given by the last term of Eq.  (\ref{Frequency0}) introduces an important difference. For a value, $k_{m} >  k > 0$, where $k_{m}^2\equiv \frac{4}{27 \hbar^2 \Lambda^2}$, there are two transition hypersurfaces from the Euclidean to the Lorentzian region, $\Sigma' \equiv \Sigma(a_+)$ and $\Sigma'' \equiv \Sigma(a_-)$, respectively, with  
\begin{eqnarray}\label{amas}
a_+ & \equiv & \frac{1}{\sqrt{3 \Lambda}} \sqrt{1 + 2 \cos\left(\frac{\theta_k}{3}\right)} , \\ \label{amenos}
a_- & \equiv & \frac{1}{\sqrt{3 \Lambda}} \sqrt{1 - 2 \cos\left(\frac{\theta_k+\pi}{3}\right) } , 
\end{eqnarray}
where, in units for which $\hbar=1$,
\begin{equation}\label{thetak}
\theta_k \equiv \arctan\frac{2 k \sqrt{k_{m}^2 - k^2 } }{k_{m}^2 - 2 k^2} .
\end{equation}
The picture is then rather different from the one depicted in Fig. \ref{DS_instanton}. First, at the transition hypersurface $\Sigma'$ the universe finds the Euclidean region (let us notice that for $k\rightarrow 0$, $a_+ \rightarrow \frac{1}{\sqrt{\Lambda}}$ and $a_-\rightarrow 0$). However, before reaching the collapse, the Euclidean instanton finds the  transition hypersurface $\Sigma''$  (see Fig \ref{EE_instanton}). Then, following a mechanism that parallels that proposed by Barvinsky and Kamenshchik in Refs. \cite{Barvinsky2006, Barvinsky2007a, Barvinsky2007b}, two instantons can be matched by identifying their hypersurfaces $\Sigma''$ (see Fig. \ref{EntangledInstanton}). The instantons can thus be created in pairs which would eventually give rise to an entangled pair of universes. Let us notice that this is a quantum effect having no classical analog because the quantum correction term in Eq. (\ref{Frequency0}) does not appear in the classical theory. 

Furthermore, let us also note that there is no Euclidean regime for the value $k \geq k_m$ and, therefore, no universes are created from the space-time foam with such values of the mode. The value  $k_m$ can then be considered the natural cut-off of the theory, avoiding the ultraviolet divergences of the scalar field. The same value can also be taken as a cut-off for more general models of the minisuperspace provided that the general boundary condition for the potential of the scalar field \cite{Kiefer2005, Kiefer2007}, $V(a,\varphi)\rightarrow 0$ for $a\rightarrow 0$, is satisfied. In that case, the behavior of the modes of the universe in the limit of small values of the scale factor would be similar to that given by Eq. (\ref{equationModes}).

For a value of the scale factor $a\gg a_-$, the quantum correction term in Eq. (\ref{Frequency0}) can be disregarded and the WKB approximation can be considered. Then, the solutions of Eq. (\ref{eq01b}) are given, up to order $\hbar$, by
\begin{equation}\label{modes}
A_k(a) \approx \frac{1}{ \sqrt{2 a \, \omega(a)}} e^{\pm \frac{i}{\hbar} S_c(a)} ,
\end{equation}
where $S_c(a)$ is the solution of the corresponding Hamilton-Jacobi equation, given by
\begin{equation}\label{Caction}
S_c(a) = \int^{a} d a' \, \omega(a') = \frac{(a^2 \Lambda - 1)^\frac{3}{2}}{3 \Lambda} .
\end{equation}
The negative sign in Eq. (\ref{modes}) corresponds to the expanding branch of the universe and the positive sign to the contracting branch. This can easily be seen by noticing that in the semiclassical approximation the momentum operator, defined by the equation $\hat{p}_a \phi(a) \equiv - i \hbar \frac{\partial\phi(a)}{\partial a}$, is highly picked around the value of the classical momentum \cite{Halliwell1987}, $p_a^c \equiv - a \frac{\partial a}{\partial t}$, where $t$ is the Friedmann time. Then, $\frac{\partial a}{\partial t} \approx \mp \frac{1}{a} \frac{\partial S_c}{\partial a}$ in the semiclassical regime, where the negative sign corresponds to the positive sign in Eq. (\ref{modes}), and the positive sign corresponds to the negative one. The WKB approximation given by Eq. (\ref{modes}) is still valid in the Euclidean regime provided that $a_+ > a \gg a_-$. Then, the oscillatory function turns out to be an exponential function,  $\phi \propto e^{\pm\frac{I_E}{\hbar}}$, with $I_E(a) = iS_c(a)$. For a value, $a_+ \gg a \gtrsim a_-$, the first term in the integrand of Eq. (\ref{Frequency0}) can be disregarded and Eq. (\ref{equationModes}) can be solved in terms of modified Bessel functions, so an approximate solution can be obtained for the whole Euclidean domain (with $a_- \gg l_P$).

The kind of universes created and annihilated by the operators $\hat{c}_k^\dag$ and $\hat{c}_k$, respectively, in Eq. (\ref{decomposition}) depends on the boundary condition that is imposed on the state of single universes. The no-boundary condition \cite{Hawking1982, Hawking1983} is equivalent in the minisuperspace to imposing regularity conditions to the wave function of the universe in the Euclidean regime \cite{Halliwell1990}. Then, the negative sign has to be chosen in the Euclidean version of the modes given by Eq. (\ref{modes}). By applying appropriate matching conditions, the wave function turns out to be given in the Lorentzian regime by an equally probable combination of expanding and contracting branches of the universe \cite{Hawking1985}. Then, an effectively interaction process makes the quantum superposition to decoherence into the single branches \cite{Halliwell1989, Kiefer1992, RP2011}. It has to be noticed, however, that such a regularity condition has only to be imposed on the zero mode wave function because, for other modes different from zero, the quantum correction term in Eq. (\ref{Frequency0}) makes the Euclidean wave function to be regular, for both signs of the exponential, in the Euclidean domain defined by, $a_+^E > a^E > a_-^E \neq 0$.

On the other hand, the tunneling boundary condition \cite{Vilenkin1982, Vilenkin1986} states that the only modes that survive the quantum barrier of the Euclidean region are the outgoing modes of the minisuperspace that correspond to the expanding branches of the universe. Then, the negative sign has to be chosen in Eq. (\ref{modes}), and the modes $A_k(a)$ describe expanding branches of the universe. Thus, the operators $\hat{c}_k^\dag$ and $\hat{c}_k$ in Eq. (\ref{decomposition}) correspond to the creation and annihilation operators, respectively, of expanding branches of the universe. The modes of the wave function (\ref{decomposition}), $\phi_k(a, \varphi) = e^{i k \varphi} A_k(a)$, with the choice of the tunneling boundary condition, satisfy the orthonormality conditions
\begin{equation}\label{eq08}
(\phi_k , \phi_l) = \delta_{kl} \; , \; (\phi_k^* , \phi^*_l) = - \delta_{k l} \; , \; (\phi_k , \phi_l^*) = 0 ,
\end{equation}
under the following scalar product,
\begin{equation}\label{eq09}
(\phi, \psi) \equiv - i \int_{-\infty}^\infty d\varphi \, W^{-1} ( \phi \partial_a \psi^* - \psi^* \partial \phi ) ,
\end{equation}
where $W=\frac{1}{a}$ is the Wronskian of Eq. (\ref{equationModes}).

The universes of the multiverse can generally posses different values of the scale factor as well as different values of their cosmological constants and matter fields. The general quantum state of the multiverse would then be given by a linear combination of product states of the form \cite{RP2010}
\begin{equation}\label{stateMultiverse}
\Psi_{\vec{N}}(a, \vec{\phi}) = \Psi_{N_1}^{\Lambda_1}(a, \phi_1)  \Psi_{N_2}^{\Lambda_2}(a, \phi_2) \cdots  \Psi_{N_n}^{\Lambda_n}(a, \phi_n) ,
\end{equation}
where, $\vec{\phi}\equiv (\phi_1, \phi_2, \ldots, \phi_n)$, and $\vec{N}\equiv (N_1, N_2, \ldots, N_n)$, with $N_i$ being the number of universes of type $i$, represented by the wave function $\phi_i \equiv \phi(a, \varphi_i)$ that corresponds to a universe with the value of the cosmological constant $\Lambda_i$. The functions, $\Psi_{N_i}^{\Lambda_i}(a, \phi_i)$ in Eq. (\ref{stateMultiverse}), are the wave functions of the number eigenstates of the third quantized Schr\"{o}dinger equation
\begin{equation}\label{3Schrodinger}
i \hbar \frac{\partial}{\partial a} \Psi_{N_i}^{\Lambda_i}(a, \phi_i) = \hat{H}_i(a,\phi, p_\phi) \Psi_{N_i}^{\Lambda_i}(a, \phi_i) ,
\end{equation}
where, $\hat{H}_i(a,\phi, p_\phi) = \int d k \, \hat{H}_k^{(i)}(a,\phi_k, p_{\phi_k})$, is the third quantized Hamiltonian that corresponds to each kind of universe [see, Eq. (\ref{HamiltonianTotalMultiverse})]. The customary interpretation of the wave function of the multiverse (\ref{stateMultiverse}) is the following \cite{Strominger1990}: let us consider the expansion of the wave function in the orthonormal basis of number states, i.e.
\begin{equation}\label{stateTotal}
\Psi = \sum_{\vec{N}} \Psi_{\vec{N}}(a, \vec{\phi}) | \vec{N} \rangle ,
\end{equation}
then, $\Psi_{\vec{N}}(a_0, \vec{\phi})$ is the probability amplitude to find $\vec{N}$ universes in the state of the multiverse with a value of the scale factor $a=a_0$. The state given by Eq. (\ref{stateTotal}) only represents the quantum state of a multiverse made up of homogeneous and isotropic universes. However, the homogeneity and isotopy of the universes of the multiverse are conditions that can be assumed in a first approximation provided that the transition hypersurface $\Sigma''$ is located well above the Planck scale, i.e. $a_-\gg l_P$. Then, it can be considered that the quantum state (\ref{stateTotal}) rather generally represents the quantum state of the multiverse.

Let us finally analyze the case of a quantum multiverse made up of flat universes. Then,  the second term in the radicand of Eq. (\ref{Frequency0}) disappears and the wave function of the universe  represents the quantum state of a flat DeSitter universe with the value $\Lambda$ of the cosmological constant or, in the case of considering Eq. (\ref{eq01})  instead of Eq. (\ref{eq01b}), it represents the quantum state of a flat universe homogeneously and isotropically  filled with a  fluid with equation of state given  by $p = w \rho$, where $p$ and $\rho$ are the pressure and the energy density of the fluid, respectively. In such a case, the Wheeler-DeWitt can be solved exactly and the modes of the wave function of the universe, normalized with the tunneling boundary condition, are given by
\begin{equation}\label{eq04}
B_k(a) \equiv A_k^{{\rm flat}}(a)= \sqrt{\frac{\pi}{4 q}} e^{\frac{\pi k}{2 q}} e^{i k \varphi} \mathcal{H}_{\frac{i k}{q}}^{(2)}(\frac{\omega_0}{q \hbar} a^q) ,
\end{equation}
where $\mathcal{H}_{\nu}^{(2)}(x)$ is the Hankel function of second kind and order $\nu$. The modes given in Eq. (\ref{eq04}) correspond to the expanding branches of the universe because for large values of the scale factor, or in the limit $\hbar \rightarrow 0$, they can be written as \cite{Abramovitz1972}
\begin{equation}\label{eq05}
\mathcal{H}_\nu^{(2)}(\frac{\omega_0}{q\hbar} a^q) \sim a^{-\frac{q}{2}} e^{- \frac{i}{\hbar} S_c(a) }  ,
\end{equation}
with, $S_c(a) = \frac{\omega_0}{q} a^q$, being the classical action (let us recall that for a vacuum energy dominated universe, $q=3$ and $\omega_0\equiv\sqrt{\Lambda}$). If we otherwise impose the 'no-boundary' condition by requiring regularity conditions in the whole domain of the wave function, the modes turn out to be given by
\begin{equation}
\bar{B}_k = \left( \frac{2 q}{\pi} \sinh \frac{k \pi}{q}  \right)^{-\frac{1}{2}} e^{i k \varphi} \mathcal{J}_{-\frac{i k}{q}} (\tilde{\omega}_0 a^q) .
\end{equation}
The two sets of modes, obtained respectively with the tunneling boundary condition and the no-boundary proposal, are then related by the Bogoliubov relation
\begin{equation}\label{eq10}
\bar{B}_k = \alpha_k B_k + \beta_k B^*_k ,
\end{equation}
where $\alpha_k$ and $\beta_k$ are given by
\begin{equation}\label{eq11}
\alpha_k = e^{\frac{\pi k}{q}} \beta_k \;\; , \;\; \beta_k = \left( \frac{e^{-\frac{\pi k}{q} }}{2 \sinh \frac{\pi k}{q}} \right)^{\frac{1}{2}} .
\end{equation}
Therefore, the assumption of one or the other  boundary condition for the wave function of the universe might have distinguishable consequences within the framework of a third quantization formalism. The vacuum state that satisfies the tunneling condition, $| 0_k \rangle$, and the corresponding one of the no-boundary proposal, $|\bar{0}_k\rangle$, are defined by
\begin{equation}\label{eq12}
\hat{b}_k |0_k \rangle = 0 \;\; , \; \; \hat{\bar{b}}_k |\bar{0}_k \rangle = 0 ,
\end{equation}
where $\hat{b}_k$ and $\hat{\bar{b}}_k$ are the annihilation operators of universes in the $k$ mode  of each representation. The number operator,  $\hat{\bar{N}}_k\equiv \hat{\bar{b}}_k^\dag \hat{\bar{b}}_k$, of the modes obtained with the no-boundary condition can be computed in the vacuum of the representation given by the tunneling condition, i.e. 
\begin{equation}\label{eq13}
\langle 0_k | \hat{\bar{N}}_k | 0_k \rangle = |\beta_k|^2 = \frac{1}{e^{\frac{2 \pi  k}{q} } - 1} .
\end{equation}
It corresponds to a thermal distribution with a temperature given by
\begin{equation}\label{eq14}
T \equiv \frac{q}{2 \pi} \left( \frac{\hbar}{k_B}\right),
\end{equation}
with $q \equiv \frac{3}{2}(1-w) = 3$ for a vacuum dominated universe (for which $w=-1$). The above thermal distribution is formally similar to the thermal radiation that appears in the quantization of a scalar field in a Milne universe  in the context of the quantum field theory in a curved space-time (see Ref. \cite{Birrell1982}). However, unlike the Milne universe, it is not clear in the case of a multiverse made up of parent universes which vacuum state corresponds to a 'preferred observer' (i.e., to an adiabatic vacuum), because the modes of the wave function of the universe are defined in the minisuperspace rather than upon the space-time.

Furthermore, in the case of phantom dominated universe it appears a multiverse as a consequence of the discretization of the parameter $w$ in the equation of state of the phantom fluid  \cite{PFGD2007}. In that case, $w \rightarrow w_j\equiv -1 - \frac{1}{3 j}$ for $j=1,2,\ldots$, and then each universe of that multiverse has a different temperature given by,
\begin{equation}\label{eq17}
T_j \equiv \frac{3}{2 \pi}  (1 + \frac{1}{6 j} ) \; ; \; \; j= 1,2,\ldots, \infty .
\end{equation}
The temperature for a vacuum dominated universe, for which $w=-1$ ($j\rightarrow \infty$), presents the limiting value $T = \frac{3}{2 \pi}$. Thus, the temperature of a phantom universe is higher than that of a vacuum dominated universe.

\subsection{Entangled states}

Entangled states and other quantum states with no classical analog can generally be posed in the quantum multiverse. Let us first consider a pair of universes created from the double instantons described in the preceding section. The matching hypersurface $\Sigma'' \equiv \Sigma''(a_-)$, where $a_-\equiv a_-(\theta_k)$ is given by Eqs. (\ref{amenos}-\ref{thetak}),  depends on the value of the mode, $k$.  Therefore, the matched instantons can only be joined for an equal value of the mode of their respective scalar fields. The pair of universes created from such a double instanton are then entangled, with a composite quantum state given by
\begin{widetext}
\begin{equation}\label{EntInstanton}
\phi_{I,II} = \int dk e^{ i k (\varphi_I +\varphi_{II})} A_{I,k} (a) A_{II,k}(a) \, \hat{c}_{I,k}^\dag \hat{c}_{II,k}^\dag + e^{ - i k (\varphi_I +\varphi_{II})} A_{I,k}^*(a) A_{II,k}^*(a) \, \hat{c}_{I,k} \; \hat{c}_{II,k} ,
\end{equation}
\end{widetext}
where $\varphi_{I,II}$ are the values of the scalar field of each single universe, labelled $I$ and $II$, respectively. The cross terms like $A_{I,k}  A^*_{II,k}$ cannot be present in the state of the pair of universes because the orthonormality relations (\ref{eq08}). Then, the composite quantum state must necessarily be the entangled state represented by Eq. (\ref{EntInstanton}).

Another example of entangled states in the multiverse can be found in Ref. \cite{PFGD2007}. In a universe whose evolution is mainly dominated by the so-called phantom energy \cite{Caldwell2002} a big rip singularity occurs in our future \cite{Caldwell2002, Caldwell2009}, splitting the whole space-time manifold into two regions, before and after the big rip, respectively. These two regions are causally disconnected because of the breaking down of the classical laws of physics in the singularity. Then, we should only consider as physically admissible the space-time region before the big rip. However, the quantum effects that would dominate in the neighborhood of the big rip can smooth the singularity out \cite{Nojiri2004} and, thus, we have  quantum mechanically to consider the region after the singularity, too. Furthermore, wormholes and ringholes would crop up and grow in the neighborhood of the big rip due to the exotic nature of the phantom fluid \cite{PFGD2003, Lobo2005}, and they would eventually connect the two regions leaving the singularity outside the trajectories followed by physical signalling and making therefore accessible to any observers the contracting region beyond the big rip \cite{PFGD2007}. In both cases, however, classical causality between the regions before and after the big rip cannot be defined. In the former case, because there is no well-posed notion of causality in the quantum realm. In the latter, because the relative motion of the mouths of the wormholes makes the space-time variables of the two regions be classically uncorrelated and independent. Nevertheless, it has to be notice that the kind of fluid that drives the evolution of the space-time of both regions is the same from the very definition of the model.

Not all the values of the general solution of the Friedmann equation of the phantom multiverse are admissible, and the parameter of the equation of state, $w$, has to be discretized to obtain consistent values of the scale factor in the region beyond the big rip  \cite{PFGD2007}. Then, each value of $j$ of the discretization, $w\equiv w_j = - 1 -\frac{1}{3 j}$, can be interpreted as representative of a proper realization of the universe in the phantom multiverse, whose quantum state is then given by Eq. (\ref{decomposition}) for each value of the index $j$, i.e
\begin{equation}\label{decomposition2}
\hat{\phi}_j(a,\varphi) = \int d k e^{i k \varphi} A_{j,k}(a) \hat{c}_{j,k}^\dag + e^{-i k\varphi} A^*_{j,k}(a) \hat{c}_{j,k} ,
\end{equation}
where $A_{j,k}$ corresponds to the expanding semiclassical branch of the universe before the big rip and $A_{j,k}^*$ corresponds to the contracting semiclassical branch after the singularity. In that case, let us consider two of these realizations, determined by the quantum numbers $j$ and $l$, respectively. The quantum state of the multiverse is given, for a single mode $k$, by
\begin{equation}\label{eq18}
\phi_{jl} =  A_ {j,k}  A_{l,k} \hat{c}^\dag_{j,k} \hat{c}^\dag_{l,k} +  A_ {j,k} ^{*} A_{l,k}^{*} \hat{c}_ {j,k}  \hat{c}_{l,k} .
\end{equation}
Taking into account once again the orthogonality relations between the modes of the universe, given by Eq.  (\ref{eq08}), there cannot be terms like $ A_ {j,k}A_{l,k}^{*}$ , and therefore the quantum state of the universe must be the entangled state given by Eq. (\ref{eq18}). 

Entangled states have no classical analog and they provide us with an example in which the EPR argument could be applied in a cosmological context. However, there is no need of a common space-time among the universes in the quantum multiverse and, therefore, the concepts of locality and non-locality would not longer be applied. The entangled states in the quantum multiverse are rather related to the quantum interdependence of the states that represent different regions or branches of the universe, being these however classically disconnected for the reasons previously argued in this section.

\subsection{Squeezed states}

In Ref. \cite{RP2010}, it is shown that the  quantum state of a multiverse made up of homogeneous and isotropic space-times is given by a squeezed state. In quantum optics, squeezed states  are considered quantum states with no classical analog and they are related with the violation of  classical inequalities. We shall first obtain the quantum state of the multiverse in terms of squeezed states and, in the next subsection, we shall analyze the violation of classical inequalities as well as the EPR argument in the context of a quantum multiverse.

Let us first notice that Eq. (\ref{equationModes}) for the modes $A_k(a)$ of the wave function of the universe can formally be viewed as the equation of a damped harmonic oscillator with a \emph{time} dependent frequency given by Eq. (\ref{Frequency0}), where the scale factor formally plays the role of the time variable of the minisuperspace parametrized by the variables $(a,\varphi)$. It is well known that the Hamiltonian of a harmonic oscillator with a time dependent frequency is not an invariant operator \cite{Lewis1969}, and that its eigenstates evolve as squeezed states \cite{Lewis1969, Pedrosa1987, Dantas1992, Sheng1995, Song2000, Park2004, Vergel2009}. It implies that the representation defined by the eigenstates of the Harmonic oscillator is not a proper representation for the state of the multiverse because, then, the number of universes of the multiverse would depend on the value of the scale factor of a particular single universe. Similarly, the representation chosen in Sec. II.A in terms of the operators $\hat{c}_k$ and $\hat{c}_k^\dag$, defined after Eq. (\ref{decomposition}), is not  a proper representation for the quantum state of the multiverse because the eigenvalues of the number operator, $\hat{N}_k \equiv \hat{c}_k^\dag \hat{c}_k$, are not scale factor invariant either, a property which is not expected in the multiverse.

The boundary condition of the multiverse that the number of universes does not depend on the value of the scale factor of a particular single universe determines the representation that has to be chosen. This is given by the Lewis representation  \cite{Lewis1969, RP2010}, defined for the $k$ mode by the following annihilation and creation operators  \cite{Lewis1969, RP2010}
\begin{eqnarray}\label{eq20}
\hat{b}_k(a) & = & \sqrt{\frac{1}{2 \hbar}} \left(\frac{\hat{\phi}_k}{R_k} + i (R_k  \hat{p}_{\phi_k} - \dot{R}_k \hat{\phi}_k ) \right) , \\  \label{eq21}
\hat{b}_k^\dag(a) & = & \sqrt{\frac{1}{2 \hbar}} \left(\frac{\hat{\phi}_k}{R_k} - i (R_k \hat{p}_{\phi_k}- \dot{R}_k \hat{\phi}_k ) \right) ,
\end{eqnarray}
where $R_k\equiv R_k(a)$ is a real and a non-degenerating  solution of the auxiliary equation \cite{Lewis1969},
\begin{equation}\label{eq22}
\ddot{R}_k + \frac{1}{a} \dot{R}_k + \frac{\omega^2_k(a)}{\hbar^2} R_k = \frac{1}{a^2 R_k^3}  .
\end{equation}
It can be checked that a solution of Eq. (\ref{eq22}) is given by, $R_k = \sqrt{A_{1,k}^2 + A_{2,k}^2}$, where $A_{1,k}$ and $A_{2,k}$ are two linearly independent solutions of Eq. (\ref{equationModes}) satisfying the normalization condition $A_{1,k} \dot{A}_{2,k} - A_{2,k} \dot{A}_{1,k} =\frac{1}{a}$. In the following the label $k$ will be omitted assuming that we are always dealing with a particular single mode of the wave function of the universe except otherwise indicated.

The operators defined in Eqs. (\ref{eq20}) and (\ref{eq21}) satisfy the usual relations,
\begin{eqnarray} \label{eq23}
\hat{b}(a) | N, a \rangle &=& \sqrt{N} | N - 1, a \rangle , \\ \label{eq24} \hat{b}^\dag(a) | N , a \rangle &=& \sqrt{N+1} | N +1, a \rangle , \\ \label{eq25}
\hat{b}^\dag(a) \hat{b}(a) | N ,a \rangle &=& N | N ,a \rangle ,
\end{eqnarray}
where  $| N, a \rangle$ are the eigenstates of the invariant operator, $\hat{I} \equiv \hat{b}^\dag(a) \hat{b}(a) +\frac{1}{2}$, and therefore $N\neq N(a)$. Thus, $N$ can be interpreted as the number of universes in the multiverse, and $\hat{b}^\dag(a)$ and $\hat{b}(a)$ as the creation and annihilation operators of universes, respectively.

The creation and annihilation operators defined by Eqs. (\ref{eq20}-\ref{eq21}) can be related to the creation and annihilation operators of  the harmonic oscillator with constant frequency $\omega_0$, $\hat{c}^\dag$ and $\hat{c}$, respectively, by the squeezed transformation 
\begin{eqnarray} \label{eq26}
\hat{b}(a) &=& \mu_0 \,  \hat{c} + \nu_0 \,  \hat{c}^\dag , \\ \label{eq27} \hat{b}^\dag(a) &=& \mu_0^*  \, \hat{c}^\dag + \nu_0^*  \, \hat{c} ,
\end{eqnarray}
where
\begin{eqnarray}\label{eq28}
\mu_0 &=& \frac{1}{2\sqrt{\omega_0}} \left(  \frac{1}{R} + \omega_0 R - i \dot{R} \right) , \\  \label{eq29} \nu_0 &=& \frac{1}{2\sqrt{\omega_0}} \left( \frac{1}{R} - \omega_0R - i \dot{R} \right),
\end{eqnarray}
with, $|\mu_0|^2 - |\nu_0|^2 = 1$.

In quantum optics, the squeezed states of light are also dubbed two-photon coherent states \cite{Yuen1975, Yuen1976} because they can be interpreted as the coherent state of an entangled pair of photons. It allows us to interpret the squeezed states of the multiverse as the state of a  correlated pair of universes. Let us notice that in the Lewis representation the Hamiltonian for each single mode of the state of the multiverse, $\hat{H}=\frac{1}{2}\hat{p}_\phi^2 +\frac{\omega^2}{2}\hat{\phi}^2$, turns out to be \cite{RP2009, RP2010}
\begin{equation}\label{eq31}
\hat{H} = \hbar \left[ \beta_- \hat{b}^2 + \beta_+(\hat{b}^\dag)^2 + \beta_0 \left( \hat{b}^\dag \hat{b} + \frac{1}{2} \right) \right] ,
\end{equation}
with
\begin{eqnarray}\label{eq32}
\beta_+^* = \beta_- &=& \frac{1}{4} \left\{ \left( \dot{R} - \frac{i}{R} \right)^2 + \omega^2 R^2  \right\} , \\ \label{eq33}
\beta_0 &=& \frac{1}{2} \left( \dot{R}^2 + \frac{1}{R^2} + \omega^2 R^2 \right) .
\end{eqnarray}
It is worthy to note that the Hamiltonian given by Eq. (\ref{eq31}) is formally equivalent to the Hamiltonian of the degenerated parametric amplifier of quantum optics, which is associated to the creation and annihilation of pairs of photons. Similarly, the quadratic terms of $\hat{b}^\dag$ and $\hat{b}$ in the Hamiltonian (\ref{eq31}) can be associated to the creation and annihilation of correlated pairs of universes in the quantum state of the multiverse.

We can  define the creation and annihilation operators of pairs of degenerated universes, i.e. those with the same properties and $\hat{b}_1\equiv \hat{b}_2$, $\hat{B}$ and $\hat{B}^\dag$,  as
\begin{eqnarray}\label{eq34}
\hat{B}(a) &=& \cosh r \; \hat{b} + e^{-i\frac{\theta}{2}} \sinh r \;  \hat{b}^\dag , \\ \label{eq35} \hat{B}^\dag(a) &=& \cosh r \;  \hat{b}^\dag +  e^{i\frac{\theta}{2}} \sinh r \;  \hat{b} ,
\end{eqnarray}
where
\begin{eqnarray}
\sinh 2r &=& \frac{2 |\beta_\pm |}{\omega} , \\ \cosh 2r &=& \frac{\beta_0}{\omega} , \\ \theta &=& i \ln\frac{\beta_+}{\beta_-} ,
\end{eqnarray}
with $\beta_\pm$ and $\beta_0$ being  defined in  Eqs. (\ref{eq32}-\ref{eq33}). In terms of the creation and annihilation operators of correlated pairs of universes the Hamiltonian recovers the diagonal representation, i.e.  $\hat{H} = \hbar \omega (\hat{B}^\dag \hat{B} + \frac{1}{2})$.  Thus, it can be interpreted that the quantum correlations between the states of the multiverse, which are given by the non-diagonal terms in the Hamiltonian, disappear when the universes are considered in pairs. However, a pair of universes forms a entangled state for which the thermodynamic properties of entanglement of each individual universe can be computed.

Let us now define two other representations  with a clear physical interpretation of the state of the multiverse. We can consider large universes with a characteristic length of order of the Hubble length of our universe. They are called parent universes  \cite{Strominger1990}. For large values of the scale factor,  the non-diagonal terms in the Hamiltonian (\ref{eq31}) vanish and the coefficient $\beta_0$ asymptotically coincide with the proper frequency of the Hamiltonian \cite{RP2010}. Equivalently, it can be checked that  $r \rightarrow 0$ in Eqs. (\ref{eq34}-\ref{eq35}) and, therefore, the operators $\hat{B}^\dag$ and $\hat{B}$ turn out to be the creation and annihilation operators of single universes. Then, the quantum correlations between the number states disappear and, thus, the quantum transitions among number states are asymptotically suppressed for parent universes. In terms of the creation and annihilation of parent universes, asymptotically defined by $\hat{b}_p^\dag  \equiv \sqrt{\frac{\omega(a)}{ 2 \hbar}} ( \hat{\phi} - \frac{i}{\omega(a)}  \hat{p}_\phi )$ and $\hat{b}_p \equiv  \sqrt{\frac{\omega(a)}{ 2 \hbar}} ( \hat{\phi} + \frac{i}{\omega(a)}  \hat{p}_\phi )$, the creation and annihilation operators of Lewis states, given by Eqs. (\ref{eq20}-\ref{eq21}), are
\begin{eqnarray}\label{eq39}
\hat{b}(a) &=& \mu_p \, \hat{b}_p + \nu_p  \, \hat{b}_p^\dag , \\ \label{eq40}
\hat{b}^\dag(a) &=& \mu_p^* \,  \hat{b}_p^\dag + \nu_p^*  \, \hat{b}_p ,
\end{eqnarray}
where
\begin{eqnarray}
\mu_p &=& \frac{1}{2 \sqrt{\omega(a)}} (\frac{1}{R} + \omega(a) R - i \dot{R} ) , \\
\nu_p &=& \frac{1}{2 \sqrt{\omega(a)}} (\frac{1}{R} - \omega(a) R - i \dot{R} ) ,
\end{eqnarray}
with $|\mu_p |^2 - |\nu_p |^2 = 1$, so they are related by a squeezed transformation, too. 

We can also consider the quantum fluctuations of the space-time of a parent universe, whose contribution to the wave function of the universe is important at the Planck scale \cite{Wheeler1957}. Some of these fluctuations can be viewed as tiny regions of the space-time that branch off from the parent universe and rejoin the large regions thereafter; thus, they can be then interpreted as virtual baby universes \cite{Strominger1990}. In that case, for small values of the scale factor \cite{RP2010},
\begin{eqnarray}\label{baby01}
\beta_+^* = \beta_-  & \rightarrow & -\frac{\omega_b}{4} , \\ \label{baby02}
\beta_0 & \rightarrow & \frac{\omega_b}{2} ,
\end{eqnarray}
where $\omega_b$ is a constant that would generally depend on the length and energy scales of the baby universe. Quantum correlations play then an important role in the state of the gravitational vacuum. This is represented by a squeezed state, an effect that can be related to that previously pointed out by Grishchuck and Sidorov \cite{Grishchuk1990}, who also showed that the squeezed state of the gravitational vacuum can be interpreted as the creation of gravitational waves in an expanding universe. In terms of the creation and annihilation operators of baby universes, $\hat{b}_b^\dag  \equiv \sqrt{\frac{\omega_b}{ 2 \hbar}} ( \hat{\phi} - \frac{i}{\omega_b}  \hat{p}_\phi )$ and $\hat{b}_p \equiv  \sqrt{\frac{\omega_b}{ 2 \hbar}} ( \hat{\phi} + \frac{i}{\omega_b}  \hat{p}_\phi )$, the Lewis operators (\ref{eq20}-\ref{eq21}) are given by
\begin{eqnarray}\label{eq46}
\hat{b}(a) &=& \mu_b \, \hat{b}_b + \nu_b \,  \hat{b}_b^\dag , \\ \label{eq47}
\hat{b}^\dag(a) &=& \mu_b^*  \, \hat{b}_b^\dag + \nu_b^*  \, \hat{b}_b ,
\end{eqnarray}
with,
\begin{eqnarray}
\mu_b &=& \frac{1}{2 \sqrt{\omega_b}} (\frac{1}{R} + \omega_b R - i \dot{R} ) , \\
\nu_b &=& \frac{1}{2 \sqrt{\omega_b}} (\frac{1}{R} - \omega_b R - i \dot{R} ) ,
\end{eqnarray}
and, $|\mu_b |^2 - |\nu_b |^2 = 1$.

Let us finally consider the general quantum state of a multiverse which is made up of pairs of entangled universes. As it has been pointed out in Sect. II.A, the universes of the multiverse can generally posses different values of their cosmological constants and scalar fields. However, the value of the cosmological constant for each single pair of universes is the same as a result of the boundary condition imposed on the state of the whole multiverse. Then, the general quantum state of the multiverse would evolve then with the Schr\"{o}dinger equation (\ref{3Schrodinger}), with the Hamiltonians $\hat{H}_i$ given by
\begin{widetext}
\begin{equation}\label{HamiltonianTotalMultiverse}
\hat{H}_i(a,\phi, p_\phi) = \hbar \int d k \left\{ \beta_{k,-}^{(i)} \hat{b}_{1,k}^{(i)} \hat{b}_{2,k}^{(i)} + \beta_{k,+}^{(i)} (\hat{b}_{1,k}^{(i)})^\dag (\hat{b}_{2,k}^{(i)})^\dag  + \frac{1}{2} \beta_{0,k}^{(i)} \left( (\hat{b}_{1,k}^{(i)})^\dag \hat{b}_{1,k}^{(i)}  + (\hat{b}_{2,k}^{(i)})^\dag \hat{b}_{2,k}^{(i)} + 1\right)  \right\} ,
\end{equation}
\end{widetext}
where, $\beta_{k,\pm}^{(i)}$ and $\beta_{k,0}^{(i)}$ are given by Eqs. (\ref{eq32}-\ref{eq33}) for each value $k$ of the mode of the wave function of a single universe, and the index $i$ labels the different species of pairs of universes that can be present in the multiverse. In the case considered in Sec. II.A, it labels the different values of the cosmological constant, $\Lambda_i$, and the scalar field, $\varphi_i$, of the universes of the multiverse [see the definitions given after Eq. (\ref{stateMultiverse})].

\subsection{Violation of classical equations in the multiverse}

\begin{figure}

 \includegraphics[width=8cm]{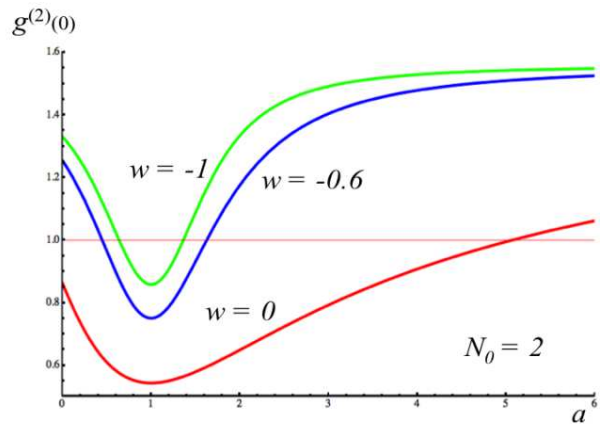}                
 \caption{Antibunching effect  in the multiverse, Eq. (\ref{eq42}), for different values of the parameter $w$ in the equation of state of the fluid that dominates the expansion of the universe.}
\label{fig:fcorrelacion}
\end{figure}

In quantum optics, squeezed states and other quantum states violate some inequalities that should be satisfied in the classical description of light. For instance, the second order coherence, $g^{(2)}(0)$, which classically should satisfy \cite{Reid1986, Scully1997} $g^{(2)}(0)\geq 1$, quantum mechanically is given, for a single mode, by \cite{Reid1986}
$$
g^{(2)}(0) = \frac{\langle (\hat{a}^\dag)^2 \hat{a}^2\rangle}{\langle \hat{a}^\dag \hat{a}\rangle^2} ,
$$
where $\hat{a}$, $\hat{a}^\dag$ are boson operators satisfying the commutation relation $[\hat{a},\hat{a}^\dag]=1$. In the quantum state of the multiverse, taking into account the relations (\ref{eq26}-\ref{eq27}) for the operators $\hat{b}$ and $\hat{b}^\dag$, the second order coherence function can be written as
\begin{equation}\label{eq42}
g^{(2)}(0) = 1 + \frac{14 x^4 + 9 x^2 - 2}{25 x^4 + 20 x^2 + 4} ,
\end{equation}
where, $x \equiv |\nu_0|$. The function (\ref{eq42}) is represented in Fig. \ref{fig:fcorrelacion} for different values of the parameter $w$ and for $N_0 =2$. For values of the scale factor which are closed to the value $a = 1$, $N_{eff} \equiv |\nu_0|^2 \approx 2$, and the second order coherence is less than one (see Fig. \ref{fig:fcorrelacion}). For smaller and larger values of the scale factor the effect disappears, which is  consistent  because the representation in terms of the modes of the harmonic oscillator with constant frequency , $|N\rangle$ ($\omega=1$), is equivalent, except for an irrelevant phase,  to the Lewis representation, $|N, a\rangle$, with a value of $a=1$. For values, $a\gg 1$ or $a \ll 1$, the effective number of universes is large and the quantum correlations disappear. It clearly reveals the strong dependence of the violation of the classical inequalities on the  representation chosen to describe the quantum state of the multiverse.

\begin{figure}
  \centering
 \includegraphics[width=9cm, height=4.5cm]{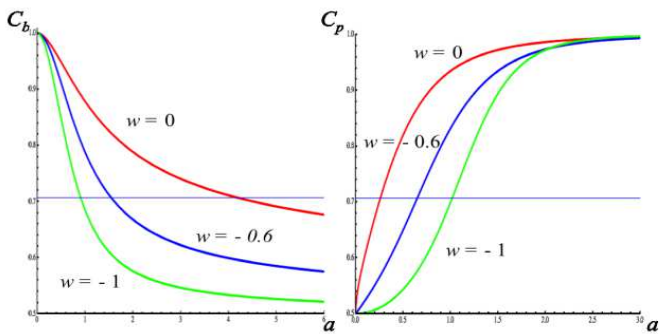}                
  \caption{Violation of Bell inequalities, Eq. (\ref{eq6222}), for different values of the parameter $w$: i) for baby universes, $C_b$, and ii) for parent universes, $C_p$.}
  \label{fig:Bell}
\end{figure}

Squeezed states violate  the Cauchy-Schwartz inequality for any value of the squeezing parameters \cite{Reid1986}, and they can also violate the Bell's inequalities. The latter violation is even more important because it is directly related to the non-local characteristic of the quantum theory. Bell's inequalities are violated, for a two mode state, when  \cite{Reid1986}
\begin{equation}\label{eq6222}
C=\frac{\langle \hat{b}_1^\dag \hat{b}_1 \hat{b}_2^\dag \hat{b}_2 \rangle}{\langle \hat{b}_1^\dag \hat{b}_1 \hat{b}_2^\dag \hat{b}_2 \rangle + \langle (\hat{b}_1^\dag)^2 \hat{b}_1^2 \rangle} \geq \frac{\sqrt{2}}{2} .
\end{equation}
In the multiverse, taking into account Eqs. (\ref{eq26}-\ref{eq27}), it is obtained
\begin{widetext}
\begin{eqnarray}\label{eqA}
\langle (\hat{b}_1^\dag)^2 \hat{b}_1^2 \rangle &=& N^2 ( 6 x^4 + 6 x^2 + 1) + N (6 x^4 + 2 x^2 -1) + 2 x^4 , \\ \label{eqB}
\langle \hat{b}_1^\dag \hat{b}_1 \hat{b}_2^\dag \hat{b}_2 \rangle &=& N^2 (6 x^4 + 6x^2 +1) + N (6 x^4 + 4 x^2) + x^2 (2 x^2 +1) ,
\end{eqnarray}
\end{widetext}
where, $x \equiv |\nu_0 | = \sinh r$, and $N_1 = N_2 \equiv N$.  In Eqs. (\ref{eqA}-\ref{eqB}), it has been considered that the universes are identical except for the existence of conscious observers that make each single universe distinguishable, so that $[\hat{b}_i, \hat{b}_j^\dag] = 0$, for $i \neq j$. For a initial vacuum state, $r=0$ and $N=0$, $C=1 > 0.7$, what implies a maximal violation of the Bell inequalities \cite{Reid1986}. For $N=1$, i.e. for a pair of entangled universes, it is obtained
\begin{equation}
C = \frac{14 x^4 + 11 x^2 + 1}{28 x^4 + 19 x^2 + 1} ,
\end{equation}
and Bell's inequalities are violated ($C \geq 0.7$) for, $0 < \sinh r < 0.31$, i.e. for small values of the squeezing parameter. In Fig. \ref{fig:Bell}, it is represented the violation of Bell inequalities for baby and parent universes with different values of the parameter $w$.

\subsection{EPR argument in the multiverse}

The existence  of squeezed and entangled states in the multiverse allows us to pose an argument analog to  the EPR argument in quantum mechanics. The original EPR argument \cite{Einstein1935} tried to show the incompleteness of the quantum theory from the point of view of a realistic theory. It was Bell \cite{Bell1987} who  pointed out that the EPR experiment showed  the non-local characteristic of quantum mechanics, actually. Roughly speaking, in a entangled state between two particles, we can know the properties of a distant particle by means of making a measurement on the other particle of the pair, no matter how far are they separated. As it is well known, it does not violate the principle of relativity because a classical channel is needed to transmit information from an emitter to a receiver, and thus information cannot travel at a speed greater than light.

Nevertheless, the effect of measuring a property in one of the particles determines the value of the same property in the other distant particle of the pair. That was dubbed \emph{spooky action} and it is fundamentally related to the non-local characteristic of quantum correlations. In the multiverse, entangled states as those given in Eq. (\ref{eq18}) can be posed. However, there is no common space-time between the universes in the quantum multiverse and, in such a context, the concepts of locality and non-locality do not make sense and have to be extended to the concepts of independence or interdependence of the quantum states of the universes. The entangled states of the multiverse are therefore  related to the \emph{non-separability}  of the states that correspond to different regions of the space-time, which are however classically and causally disconnected. Furthermore, the quantum correlations might have observable consequences in the dynamics and the thermodynamics of a single universe, one of which could well be a contribution to the vacuum energy of the universe in the form of an energy of entanglement between two universes.

Entangled states are also used in quantum optics to develop a protocol of teleportation by  which a particular quantum state is transmitted from an emitter (Alice) to a receptor (Bob). In the phantom multiverse, we can consider the entangled state between the regions before and after the big rip singularity, given by Eq. (\ref{eq18}). The modes of the scalar field which correspond to the expanding and contracting branches of the universe, before and after the singularity, respectively, are entangled and might be used as the quantum channel of the protocol of teleportation. As a classical channel one could consider the wormholes that crop up at both sides of the singularity  \cite{PFGD2003, Lobo2005} and, therefore, both channels would permit the transmission of a quantum state between Alice, who is located in the region before the singularity, and Bob, who is placed in a hypersurface of the space-time after the singularity. Evidently, this is only a hypothetic example of how the existence of entangled states in the multiverse could provide us with non-classical communication protocols between different universes by making use of the non-separable nature of their quantum states.

Let us finally make the plausible objection that the existence of entangled states in the multiverse may be viewed as an incorrect choice of the subspaces $\mathcal{H}_1$ and $\mathcal{H}_2$ of the whole Hilbert space, $\mathcal{H}$, that corresponds to the complete quantum description of the universe. That is to say, that $\mathcal{H}$ cannot be given by a direct product, i.e. $\mathcal{H} \neq \mathcal{H}_1 \otimes \mathcal{H}_2$, or that the splitting of the whole Hilbert space in two subspaces is just a useful mathematical resource   to obtain the quantum state in $\mathcal{H}$ that corresponds to a unique single universe. It can be accepted. However, the analog argument in the quantum description of the electromagnetic field would be that entangled states between a pair of photons are just a useful way to represent the state of the field. The violation of classical inequalities in quantum optics reveals however the corpuscular nature of the photon and its existence as an autonomous entity, although not necessarily independent. It allows us, in a second quantization formalism, to interpret the different modes of the wave function of the universe as different universes. The complementarity characteristic of the quantum theory impels us to consider as well their wave properties and thus quantum interference and correlations between the states of different universes, which can be considered to be identical, as in the model consider in this paper, except for the plausible existence of conscious  observers that might communicate each other through classical or quantum channels. 

In the following section we make use of this argument to compute the thermodynamics quantities of entanglement between two universes whose quantum states are correlated.

\section{Entanglement thermodynamics in the multiverse}

\subsection{Quantum thermodynamics}

For a physical system that is quantum mechanically represented by a density matrix $\hat{\rho}$, with a dynamic determined by a Hamiltonian operator $\hat{H}$, the following thermodynamical quantities can be defined \cite{Alicki2004, Gemmer2009}
\begin{eqnarray}\label{eq631}
E(a) &=& {\rm Tr}\left( \hat{\rho}(a) \hat{H}(a) \right) , \\  \label{eq632}
Q(a) &=& \int^a {\rm Tr}\left( \frac{d \hat{\rho}(a')}{d a'} \hat{H}(a') \right) da' , \\ \label{eq633}
W(a) &=& \int^a {\rm Tr}\left( \hat{\rho}(a') \frac{d \hat{H}(a')}{d a'}  \right) da' ,
\end{eqnarray}
where ${\rm Tr}(\hat{O})$ means the trace of the operator $\hat{O}$, and in the case of the multiverse the time variable has been substituted by the scale factor, which is the  temporal variable in the second quantization formalism of the minisuperspace. In these definitions, $E$ is the quantum informational analog of the energy, $Q$ is that of the heat and $W$ the analog of the work. Then, the first principle of thermodynamics, 
\begin{equation}\label{eq634}
dE = \delta W + \delta Q ,
\end{equation}
is  satisfied. The quantum entropy is defined by the von Neumann's formula,
\begin{equation}\label{eq635}
S(\hat{\rho}) = -   {\rm Tr}\left( \hat{\rho}(a) \ln \hat{\rho}(a) \right)  ,
\end{equation}
where the logarithmic function of an operator must be read as its series development, i.e.
\begin{equation}
S(\hat{\rho}) =  \sum_{k = 1}^{\infty} \frac{1}{k} \sum_{l=0}^k (-1)^l  \left( \begin{array}{c} k \\ l \end{array} \right)  {\rm Tr} \left( \hat{\rho}^{l+1} \right) = - \sum_i \lambda_i \ln \lambda_i ,
\end{equation}
being $\lambda_i$ the eigenstates of the density matrix and,  $0 \ln 0 \equiv 0$. For a pure state, $\hat{\rho}^n = \hat{\rho}$, and $\lambda_i = \delta_{i j}$ for some value $j$, and therefore the entropy vanishes.

It is worthy to note that the quantum thermodynamical energy and entropy are invariant under a unitary evolution of the state of the universe. By using the cyclic property of the trace, it can be written
\begin{eqnarray}\nonumber
S(\hat{\rho}) &=&  \sum_{k = 1}^{\infty} \frac{1}{k} \sum_{l=0}^k (-1)^l  \left( \begin{array}{c} k \\ l \end{array} \right) {\rm Tr}\left( \hat{\mathcal{U}}_S^\dag(a) \hat{\rho}_0^{l+1} \hat{\mathcal{U}}_S(a) \right) \\ \label{eq635b} &=& S(\hat{\rho}_0) ,
\end{eqnarray}
and analogously for the energy $E$,  provided that no dissipative terms or other interacting processes are considered in the dynamics of the universe. Such processes can make the state of the universe effectively undergo an non unitary evolution, and the entropy of the universe grows as the universe is expanding \cite{Joos2003, RP2011}.

The invariance given in Eq. (\ref{eq635b}) is no necessarily applicable to the heat $Q$ and work $W$. For instance, let us consider two representations, $A$ and $B$, which were related by a unitary transformation $\hat{\mathcal{U}}$, so that, $\hat{\rho}_B = \hat{\mathcal{U}} \hat{\rho}_A \hat{\mathcal{U}}^\dag$ and $\hat{H}_B = \hat{\mathcal{U}} \hat{H}_A \hat{\mathcal{U}}^\dag$. Then, it is satisfied that $E_A = E_B$ and $S_A=S_B$. In particular,
\begin{widetext}
\begin{equation}
\delta Q_A(a) + \delta W_A(a) = {\rm Tr} \left( \frac{\partial \hat{\rho}_A}{\partial a} \hat{H}_A \right) + {\rm Tr} \left( \hat{\rho}_A \frac{\partial \hat{H}_A}{\partial a} \right) = {\rm Tr} \left( \frac{\partial \hat{\rho}_B}{\partial a} \hat{H}_B \right) + {\rm Tr} \left( \hat{\rho}_B \frac{\partial \hat{H}_B}{\partial a} \right) = \delta Q_B(a) + \delta W_B(a) ,
\end{equation}
\end{widetext}
where it has been used that, $\dot{\hat{\mathcal{U}}} \hat{\mathcal{U}}^\dag = - \hat{\mathcal{U}} \dot{\hat{\mathcal{U}}}^\dag$. However, if $\hat{\mathcal{U}} \equiv \hat{\mathcal{U}}(a)$, it is not necessarily true that, $\delta Q_A = \delta Q_B$ and $\delta W_A =\delta W_B$, although the first principle of thermodynamics is still satisfied, $dE = \delta Q_A + \delta W_A = \delta Q_B + \delta W_B$. In classical thermodynamics the heat and work  are not, unlike the energy and the entropy, functions of state because their values depend on the path of integration of $\delta Q$ and $\delta W$. The analogy in quantum thermodynamics is that $Q$ and $W$ depend on the representation that is taken to compute them.

In the change of  entropy, two terms can be distinguished: one due to the variation of heat, and the change caused by the rest of processes. It can be written as \cite{Alicki2004, Kiefer2007},
\begin{equation}\label{eq637}
\frac{d S}{d a} = \frac{1}{T} \frac{\delta Q}{d a} + \sigma(a) ,
\end{equation}
where the second term, $\sigma(a)$, is called \cite{Alicki2004} \emph{production of entropy}. The second principle of thermodynamics states that the entropy of a system cannot decrease under any adiabatic process, which is equivalent to say that the production of entropy has to be always non-negative, i.e.
\begin{equation}\label{eq638}
\sigma(a) \geq 0 .
\end{equation}

It has to be noticed that in the quantum thermodynamics of open systems \cite{Kiefer2007, Alicki2004}, the change of entropy is also expressed also as $\frac{d S}{d t}= \left(\frac{d S}{d t}\right)_{{\rm ext}} + \left(\frac{d S}{d t}\right)_{{\rm int}}$, where $\left(\frac{d S}{d t}\right)_{{\rm ext}} = \frac{\delta Q}{T}$, is interpreted as the change in the entropy because the interaction with an \emph{external} bath (or reservoir) at temperature $T$; and $\left(\frac{d S}{d t}\right)_{{\rm int}} \geq 0$, is interpreted as the change of entropy because the change of the internal degrees of freedom. However, in the multiverse the terms \emph{external} and \emph{internal} lack their meaning because in a closed system all the thermodynamical magnitudes are obviously internal to the system.  We can still formally define the thermodynamical quantities given by Eqs.  (\ref{eq631}-\ref{eq633}) and (\ref{eq635}) in a similar way to which it is made in an open system, being their interpretation for a closed system like the multiverse, however, rather different. In the multiverse, the heat $Q$ and  work $W$ cannot be interpreted as ways of exchanging energy with a reservoir because, in the case being considered, there is no such reservoir. Similarly, the analog of the temperature $T$ does not represent the temperature of an external bath. \emph{All the thermodynamical magnitudes of a closed system are internal properties of the system}.

\begin{figure}
  \centering
 \includegraphics[width=9cm,height=7cm]{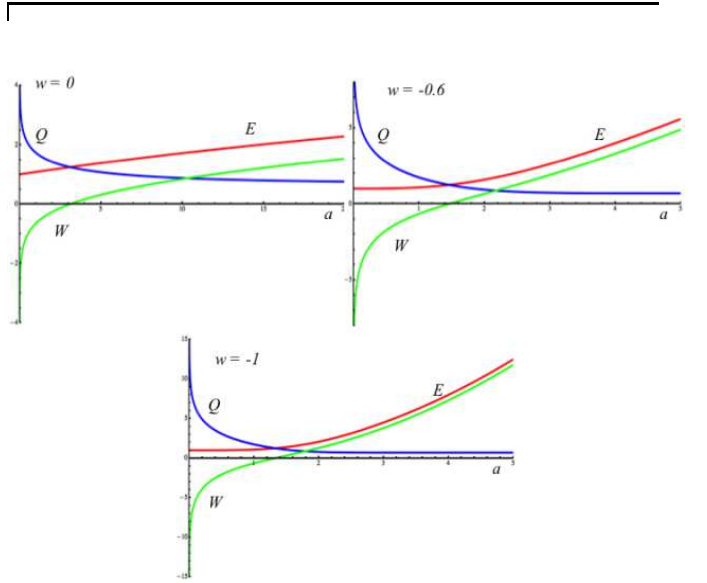}                
  \caption{Energy $E$, heat $Q$ and work $W$, Eqs. (\ref{eq6313}-\ref{eq6315}), for different values of the parameter $w$. The first principle of thermodynamics is always satisfied, $E = Q+W$.}
  \label{fig:EQW}
\end{figure}

\begin{figure}

\includegraphics[width=7cm,height=6cm]{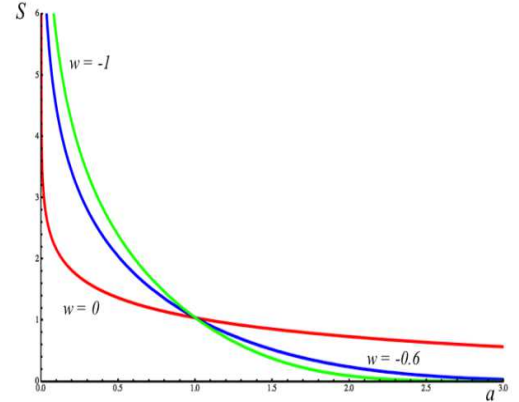}

  \caption{Quantun entropy of the universe, Eq. (\ref{eq68}), for different values of the parameter $w$. The entropy decreases with respect to the scale factor. However, the second principle of thermodynamics is still satisfied because the process is not adiabatic and the production of entropy is zero.}
  \label{fig:ES}
\end{figure}

Let us consider now the state of the multiverse and its thermodynamical properties, which depend on the boundary condition that is imposed on the state of the multiverse. Let us suppose that the boundary condition is chosen so that the multiverse is quantum mechanically represented by the  thermal state
\begin{equation}\label{eq6312}
\hat{\rho} = \frac{1}{Z} \sum_N e^{-\frac{\hbar \omega(a)}{T} (N+\frac{1}{2})} |N, a\rangle \langle N,a| ,
\end{equation}
where, $Z^{-1}\equiv 2 \sinh\frac{\hbar \omega(a)}{2 T}$, and $\omega(a) =\frac{\omega_0}{\hbar}a^{q-1}$ is the frequency of the Hamiltonian that determines the evolution of the multiverse. In that case, the thermodynamical quantities involved in the first principle of thermodynamics (\ref{eq634}) turn out to be,
\begin{eqnarray}\label{eq6313}
E(a) &=& \frac{\hbar \omega(a)}{2} \coth\frac{\hbar \omega(a)}{2 T} , \\ \label{eq6314}
Q(a) &=& T \left( \frac{\hbar \omega(a)}{2 T} \coth\frac{\hbar \omega(a)}{2 T} - \ln\sinh\frac{\hbar \omega(a)}{2 T}  \right) , \\ \label{eq6315}
W(a) &=& T \ln\sinh\frac{\hbar \omega(a)}{2 T} .
\end{eqnarray}
It  can be checked that, $E = Q+W$ and  $dE = \delta Q + \delta W$. It appears a term of heat production, $Q$, because the dependence of the frequency with respect to the scale factor. Thus, the entropy,
\begin{equation}\label{eq68}
S = \frac{\hbar \omega(a)}{2 T} \coth\frac{\hbar \omega(a)}{2 T} - \ln\sinh\frac{\hbar \omega(a)}{2 T} - \ln2 ,
\end{equation}
is no longer constant and the change of entropy,
\begin{equation}\label{eq6317}
d S = - \frac{\hbar^2 \omega \dot{\omega}}{4 T^2} \frac{1}{\sinh^2\frac{\hbar \omega(a)}{2 T}} ,
\end{equation}
turns out to be negative. However, the second principle of thermodynamics is still satisfied because the change of the entropy corresponds precisely to the change of the heat (over the temperature $T$), and the production of entropy is therefore zero,
\begin{equation}
\sigma = \frac{d S}{d a} - \frac{1}{T} \frac{\delta Q}{d a} \equiv 0.
\end{equation}
The thermodynamical magnitudes which are involved in the first principle of thermodynamics are depicted in  Fig. \ref{fig:EQW} for different values of the parameter  $w$. The entropy, depicted in  Fig. \ref{fig:ES}, decreases with the value of the scale factor although the second principle of thermodynamics is  satisfied because the process is not  adiabatic  and the production of entropy is zero, as it is expected in a closed system with no dissipative processes.

\subsection{Energy and entropy of entanglement}

The violation of classical inequalities  and the existence of entangled and squeezed states in the context of a quantum multiverse allow us to consider, in general, correlated states between two universes. It has to be noted that the entanglement depends crucially on the choice of the representation of the modes (Ref. \cite{Vedral2006}, p. 88). In this section we use two representations: the representation of parent universes that describe macroscopic universes like ours, and the representation of baby universes which can describe, in a first approximation, the quantum fluctuations of the space-time of a parent universe.

In both cases, the squeezing relations given by Eqs. (\ref{eq39}-\ref{eq39}) and (\ref{eq46}-\ref{eq47}) allow us to write the composite state of two entangled universes as,
\begin{equation}\label{eq6330}
\hat{\rho}(a) = \hat{\mathcal{U}}_S^\dag(a) | 0_1 0_2 \rangle \langle 0_1 0_2 | \hat{\mathcal{U}}_S(a) ,
\end{equation}
where the evolution operator is the squeezing operator given by,
\begin{equation}
\hat{\mathcal{U}}_S(a) = e^{r(a) e^{i \theta} \hat{b}_1 \hat{b}_2 - r(a) e^{-i \theta} \hat{b}_1^\dag \hat{b}_2^\dag} ,
\end{equation}
with $r(a)$ and $\theta(a)$ being the squeezing parameters that depend on the value of the scale factor. The boundary condition that has been taken in the composite state (\ref{eq6330}), is that the state is initially given by a pure state formed by the fundamental states of each single universe in their respective Hilbert spaces. We shall first obtain the thermodynamical properties of entanglement in terms of the squeezing parameters, $r$ and $\theta$, and  we shall then compute the value of these parameters for baby and parent universes and their thermodynamical properties of entanglement.

\begin{figure}

\includegraphics[width=6cm,height=5cm]{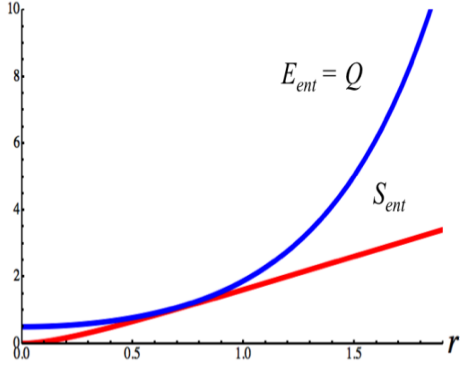}                

 \caption{Entanglement energy, integration of Eq. (\ref{eq86}) with $\omega=1$, and entanglement entropy, Eq. (\ref{eq6329}), with respect to the squeezing parameter $r$.}
 \label{fig:EQTS1}
\end{figure}

The reduced density matrix for each single universe is given by
\begin{equation}
\hat{\rho}_{(1,2)} \equiv {\rm Tr}_{(2,1)} \hat{\rho} = \sum_{N_{(2,1)}=0}^\infty \langle N_{(2,1)} | \hat{\rho} | N_{(2,1)} \rangle .
\end{equation}
Let us focus, for instance, on the universe $1$ (they both are identical anyway). Its state is given then by
\begin{equation}
\hat{\rho}_1 = \sum_{N_2=0}^\infty \langle N_2 | \hat{\mathcal{U}}_S^\dag |0_2\rangle |0_1\rangle \langle 0_1| \langle 0_2| \hat{\mathcal{U}}_S | N_2\rangle .
\end{equation}
Making use of the disentangling theorem \cite{Wodkiewicz1985, Buzek1989},  
\begin{equation}
\hat{\mathcal{U}}_S^\dag(a) = e^{\Gamma(a) e^{i \theta} \hat{b}_1^\dag \hat{b}_2^\dag} e^{- g(a) (\hat{b}_1^\dag \hat{b}_1 + \hat{b}_2^\dag \hat{b}_2 + 1)} e^{- e^{-i \theta} \Gamma(a) \hat{b}_1 \hat{b}_2} ,
\end{equation}
where,
\begin{equation}
\Gamma(a) = \tanh r(a) \;\;\; ,  \;\;\; g(a) = \ln \cosh r(a) ,
\end{equation}
we obtain that each single universe is quantum mechanically represented by the thermal state given by,
\begin{eqnarray}\nonumber
\hat{\rho}_1(a) &=& e^{- 2 g(a)} \sum_{N=0}^\infty e^{2 N \ln \Gamma(a) } |N\rangle \langle N | \\ \nonumber &=& \frac{1}{\cosh^2r} \sum_{N=0}^\infty \left( \tanh^2r\right)^N |N\rangle \langle N |  \\ \label{eq6325} &=& \frac{1}{Z} \sum_{N=0}^\infty e^{-\frac{\omega(a)}{T(a)}(N + \frac{1}{2})} |N\rangle \langle N|,
\end{eqnarray}
where, $|N\rangle \equiv |N\rangle_1$ (similarly for $\hat{\rho}_2$ with $|N\rangle \equiv |N\rangle_2$), and with,  $Z^{-1} = 2 \sinh\frac{\omega}{2 T}$. The two universes of the entangled pair evolve in thermal equilibrium with respect to each other, with a temperature that depends on the scale factor, i.e.
\begin{equation}\label{eq6327}
T \equiv T(a) = \frac{\omega(a)}{2 \ln\frac{1}{\Gamma(a)}}  .
\end{equation}
The entanglement entropy, which is defined as
\begin{equation}
S_{ent} = - {\rm Tr} (\hat{\rho}_{1} \ln \hat{\rho}_{1}) ,
\end{equation}
turns out to be 
\begin{equation}\label{eq6329}
S_{ent}(a) = \cosh^2 r \, \ln \cosh^2 r - \sinh^2 r \, \ln \sinh^2 r .
\end{equation}
It grows with respect to the squeezing parameter $r$ (Fig. \ref{fig:EQTS1}). The second principle of quantum thermodynamics, given by Eq.  (\ref{eq638}), is satisfied because the change in the entropy of entanglement corresponds precisely with the change of heat over the temperature, and the production of entropy  $\sigma$ vanishes. It can be checked by computing the thermodynamical magnitudes given by Eqs. (\ref{eq631}-\ref{eq633}).

From Eq. (\ref{eq631}), the energy of the state represented by  $\hat{\rho}_1$ ($=E(\hat{\rho}_2)$) reads
\begin{equation}
E_1(a) = {\rm Tr} \hat{\rho}_1 \hat{H}_1 = \omega (\sinh^2 r + \frac{1}{2}) = \omega (\langle \hat{N}(a) \rangle + \frac{1}{2}) ,
\end{equation}
where, $\hat{H}_1 \equiv \omega (\hat{b}_1^\dag \hat{b}_1 + \frac{1}{2})$. The change in the heat and work, given by Eqs. (\ref{eq632}) and (\ref{eq633}), respectively, are
\begin{eqnarray}
\delta W_1 &=& {\rm Tr} (\hat{\rho}_1 \frac{d \hat{H}_1}{d a} ) = \dot{\omega} (\sinh^2 r + \frac{1}{2}) , \\ \label{eq6332}
\delta Q_1 &=& {\rm Tr} (\frac{d \hat{\rho}_1}{d a} \hat{H}_1 ) = \omega \dot{r} \sinh 2r , 
\end{eqnarray}
from which it can be checked that, $dE_1 = \delta W_1 + \delta Q_1$. From Eqs. (\ref{eq6332}) and (\ref{eq6329}), it can also be checked that the production of entropy is zero, 
\begin{equation}\label{eq6333}
\sigma = \frac{d S_{ent}}{da} - \frac{1}{T} \frac{\delta Q}{d a} = 0 ,
\end{equation}
where, $T=\frac{\omega}{2} \ln^{-1}\frac{1}{\Gamma}$, is defined in Eq. (\ref{eq6327}). Moreover, Eq.  (\ref{eq6333}) can be compared with the expression which is usually used to compute the energy of entanglement (see, Refs. \cite{Mukohyama1997, Mukohyama1998, Lee2007}),
\begin{equation}
d E_{ent}  = T dS_{ent} .
\end{equation}
It enhances us to establish an energy of entanglement given by
\begin{equation}\label{eq86}
dE_{ent} = \delta Q  = \omega  \sinh 2r \; dr .
\end{equation}

\subsubsection{Parent and baby universes}

\begin{figure}
\includegraphics[width=7cm,height=6cm]{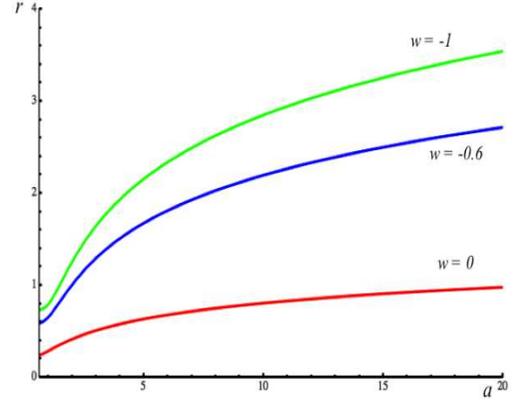}                
 \caption{Squeezing parameter, $r$, in terms of the scale factor for baby universes and for different values of the parameter $w$.}
  \label{fig:bebe1}
\end{figure}

\begin{figure}
\includegraphics[width=7cm,height=6cm]{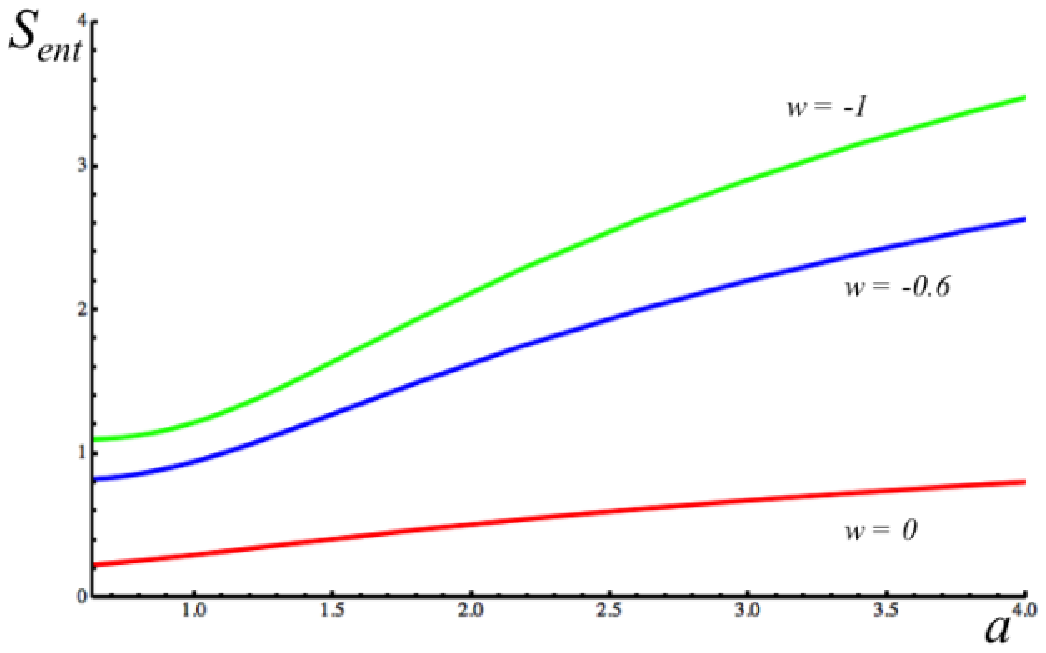}                
 \caption{Entanglement entropy in terms of the scale factor fot baby universes and for different values of the parameter $w$.}
  \label{fig:bebe2}
\end{figure}

The Lewis states are related to the states of baby universes by the squeezing relations (\ref{eq46}-\ref{eq47}). The squeezing parameter $r$ is given then by,
\begin{equation}\label{eq6341}
r = {\rm arcsinh} |\nu_b| ,
\end{equation}
The squeezing parameter, $r$, and the thermodynamical magnitudes of entanglement are depicted in  Figs. \ref{fig:bebe1}-\ref{fig:bebe2}. The work $W$ is zero because  the frequency of the harmonic oscillator that represents the baby universes, $\omega_b$ in Eqs. (\ref{baby01}-\ref{baby02}), is a constant. Then,  $E=Q$, and the total energy is the energy of entanglement. It grows with the size of the quantum fluctuations of the space-time as the universe is expanding.

\begin{figure}
\includegraphics[width=7cm,height=6cm]{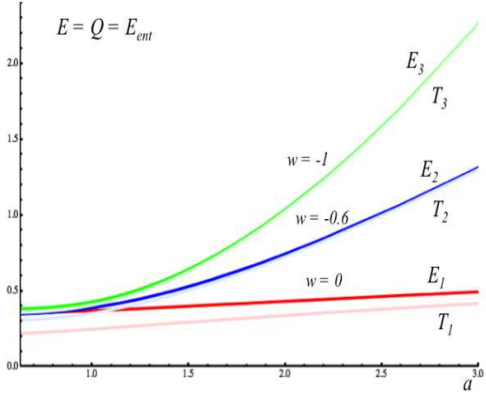}                
 \caption{Entanglement energy and temperature of entanglement of baby universes for different values of the parameter $w$.}
  \label{fig:bebe3}
\end{figure}

\begin{figure}
\includegraphics[width=7cm,height=6cm]{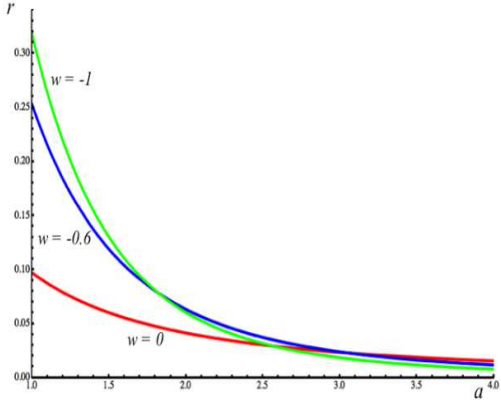}                

  \caption{Parameter of squeezing, $r$,  for parent universes and for different values of the parameter  $w$.}
  \label{fig:padre1}
\end{figure}

\begin{figure}
\includegraphics[width=7cm,height=6cm]{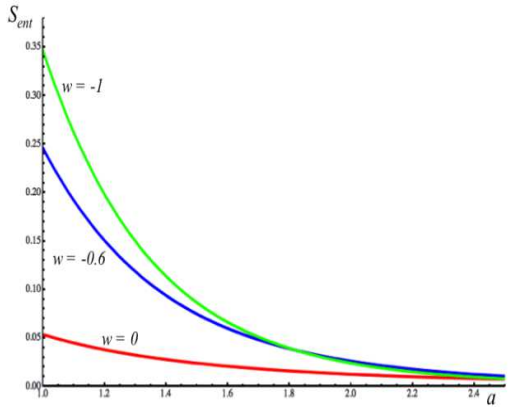}                

  \caption{Entanglement entropy in terms of the scale factor for parent universes and for different values of the parameter  $w$.}
  \label{fig:padre2}
\end{figure}

\begin{figure}
\includegraphics[width=7cm,height=6cm]{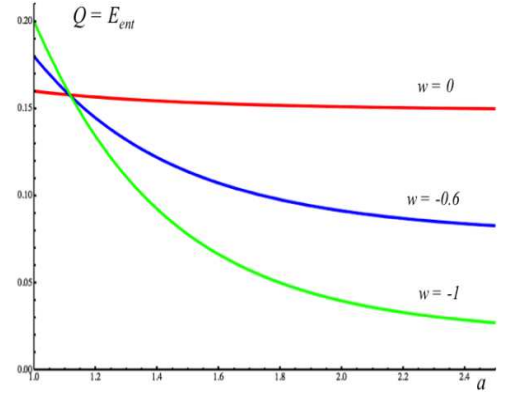}                
  \caption{Energy of entanglement for parent universes and for different values of the parameter $w$.}
  \label{fig:padre3}
\end{figure}

For parent universes, the states are given by those of a harmonic oscillator with a scale factor dependent frequency,  $\omega \equiv \omega(a) = \frac{\omega_0}{\hbar} a^{q-1}$. In that case, the Lewis states are related to the states of parent universes by the relations (\ref{eq39}-\ref{eq40}), and the squeezing parameter is
\begin{equation}\label{eq6342}
r = {\rm arcsinh} |\nu_p| .
\end{equation}
Then, the entanglement and the energy of entanglement decrease for increasing values of the scale factor. The thermodynamical properties of entanglement for parent universes are depicted in  Figs. \ref{fig:padre1}-\ref{fig:padre3}. Thus, the entanglement between two universes provides us with a mechanism by which the vacuum energy can be high in the initial stage of the universe, which is a required condition of the inflationary models, and it can however present a much smaller value in the older stages of the universe like the current one, which might fit with the current observational value of the cosmological constant.

\subsubsection{Accelerating universes}

\begin{figure}
\includegraphics[width=7cm,height=5cm]{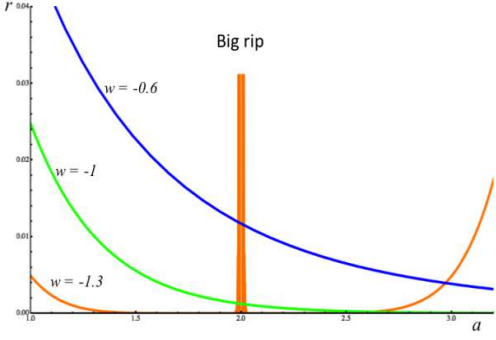}                
  \caption{Squeezing parameter for parent universes in terms of cosmic time $t$, for different values of the parameter $w$. In the big rip singularity, the thermodynamical magnitudes diverge and the model is no longer valid.}
  \label{fig:cosmico1}
\end{figure}

\begin{figure}
\includegraphics[width=7cm,height=5cm]{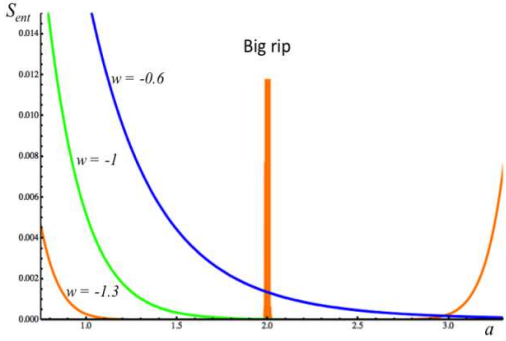}                
  \caption{Entropy of entanglement for parent universes in terms of cosmic time $t$, for different values of the parameter $w$.}
  \label{fig:cosmico2}
\end{figure}

The limit of validity of the model presented in this paper corresponds to a value of the scale factor which is well above the Planck scale. For a phantom dominated universe, moreover, the model is not valid in the achronal region around the big rip. Between those limits the semiclassical approximation is valid and we can then obtain the thermodynamical quantities of entanglement in terms of the cosmic time given by the Friedmann equation, whose solutions can be written as \cite{RP2010}
\begin{equation}\label{eq89}
a(t) = (a_0^\beta + \beta \lambda_0 (t-t_0))^{\frac{1}{\beta}} ,
\end{equation}
with, $\beta = \frac{3}{2} (1+w)$, for $w\neq -1$, and $a(t) = e^{\lambda_0 (t-t_0)}$ for $w=-1$. Inserting the value of the scale factor (\ref{eq89}) in the equations given in this section, we obtain the thermodynamical magnitudes in therms of the cosmic time (Figs. \ref{fig:cosmico1}-\ref{fig:cosmico2}, with $a_0=1$, $t_0=0$). The main feature in the case of a phantom dominated universe ($w<-1$) is the divergence of the thermodynamical magnitudes at the big rip singularity, as it was expected. The region $II$ after the singularity is symmetric to the region $I$ before the big rip. For other values of the parameter $w$, the thermodynamical properties of entanglement take the expected value for a scale factor that continuously grows in time and, therefore, with squeezing and entanglement parameters that decrease with respect to the cosmic time.

\subsubsection{Homogeneous and isotropic flat universe with a scalar field}

In Sec. II, it has been shown that the wave function of a multiverse made up of effectively flat regions of the space-time can be written in terms of two set of modes which correspond to the 'no-boundary' condition and the tunneling boundary condition, written as $\bar{B}_n$ and $B_n$, respectively. The two sets are related by the Bogoliubov transformation given in Eqs. (\ref{eq10}-\ref{eq11}), with, $|\alpha_n|^2 - |\beta_n|^2 = 1$. We can then define two different states for the vacuum of the multiverse, $| 0_n\rangle$ and $|\bar{0}_n\rangle$, respectively [see Eq. (\ref{eq12})]. Following the formal analogy with a quantum field theory in a curved space-time, the vacuum state $|\bar{0}_{k,-k}\rangle$ can be written as \cite{Mukhanov2007}
\begin{equation}\label{eq3646}
|\bar{0}_{k,-k}\rangle = \frac{1}{|\alpha_k|} \sum_{n=0}^\infty \left( \frac{\beta_k}{\alpha_k}\right)^n | n_k , n_{-k} \rangle ,
\end{equation}
where the modes $k$ and $-k$ correspond to the expanding and contracting branches of the universe, respectively. The state given by Eq.  (\ref{eq3646}) represents an entangled state between the quantum states of those branches.

Let us consider a phantom dominated universe in which the expanding and contracting branches are causally separated by the big rip singularity. In that case, the quantum state that corresponds to the regions before and after the singularity is given by a reduced density matrix which is obtained by tracing out the degrees of freedom of the complementary region. If the initial state corresponds to the vacuum state $|\bar{0}_{k,-k}\rangle$, the total density matrix is given by
\begin{eqnarray}\nonumber
\hat{\rho} &=& |\bar{0}_{k,-k}\rangle \langle \bar{0}_{k,-k} | \\ &=& \frac{1}{|\alpha_k|^2} \sum_{n,m=0}^\infty \left( \frac{\beta_k}{\alpha_k}\right)^{n+m} | n_k , n_{-k} \rangle \langle m_k , m_{-k} | .
\end{eqnarray}
The reduced density matrix for the expanding region before the singularity turns out to be
\begin{eqnarray}\nonumber
\hat{\rho}_r &=&  \frac{1}{|\alpha_k|^2} \sum_{n=0}^\infty \left( \frac{\beta_k}{\alpha_k}\right)^{2 n} | n_k  \rangle \langle n_k | \\ \label{eq80} &=& \frac{1}{\cosh^2 r_k} \sum_{n=0}^\infty \left( \tanh^2 r_k \right)^n  | n_k  \rangle \langle n_k |  .
\end{eqnarray}
It represents a thermal state with a temperature given by Eq.  (\ref{eq6327}), with $\omega_k = k$, and $\Gamma^2 = \frac{\beta_k}{\alpha_k}$, i.e.
\begin{equation}
T = \frac{k}{2 \ln\frac{|\alpha_k|}{|\beta_k|}} = \frac{q}{2 \pi} ,
\end{equation}
which is the temperature already obtained in Sec. II [see Eq.  (\ref{eq14})]. With the reduced density matrix and making use of the equations developed in this section, we can obtain the thermodynamical magnitudes that correspond to the thermal state (\ref{eq80}). The entanglement entropy, Eq. (\ref{eq6329}), turns out to be
\begin{eqnarray}\nonumber
S_{ent} &=& |\alpha_k|^2 \ln |\alpha_k|^2 - |\beta_k|^2 \ln |\beta_k|^2 \\ &=& \frac{\pi k}{q} {\rm cotanh}\frac{\pi k}{q} - \ln\left( 2 \sinh \frac{\pi k}{q} \right) ,
\end{eqnarray}
which fits with Eq. (\ref{eq68}), with $\omega_k = k$ and $T = \frac{q}{2 \pi}$. From Eqs. (\ref{eq6314}-\ref{eq6315}), it can be checked that, $Q = T S_{ent} $, and the energy and work  are given by,
\begin{eqnarray}
E &=& \frac{k}{2} {\rm cotanh}\frac{\pi k}{q} , \\
W &=& \frac{q}{2 \pi} \ln \sinh\frac{\pi k}{q} .
\end{eqnarray}

\begin{figure}

 \includegraphics[width=8cm,height=6cm]{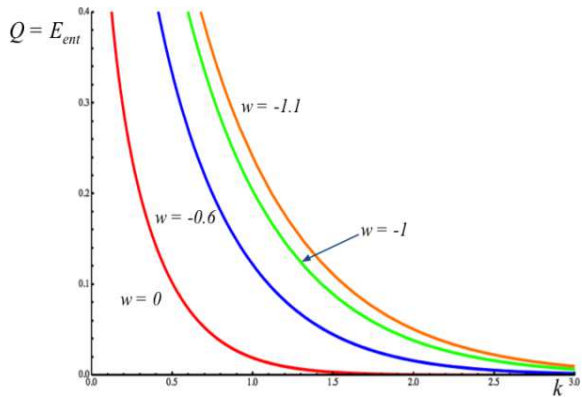}                
  \caption{Energy of entanglement with respect to the value of the mode $k$ of the scalar field, for different values of the parameter $w$ in the equation of state of the fluid that dominates the expansion of the universe.}
  \label{fig:Emodos}
\end{figure}

The change of the entropy with respect to the value of the mode, $k$,  for the constant temperature $T=\frac{q}{2 \pi}$, is [see Eq. (\ref{eq6317})],
\begin{equation}
\frac{d S}{d k} = - \frac{\pi^2 k}{q^2} \frac{1}{\sinh^2\frac{\pi k}{q}} = \frac{1}{T} \frac{\delta Q}{d k} .
\end{equation}
And therefore, the production of entropy, $\sigma$, is zero. In that case, the energy of entanglement can be identified with the heat, $Q$, i.e.
\begin{equation}
E_{ent} = Q = \frac{k}{2} {\rm cotanh}\frac{\pi k}{q} - \frac{q}{2 \pi} \ln \left( 2 \sinh \frac{\pi k}{q}\right) ,
\end{equation}
where, $q=\frac{3}{2}(1-w)$,  $w$ being the proportionality constant of the equation of state of the fluid that dominates the expansion of the universe,  $p=w \hat{\rho}$ (let us recall that $q=3$ for vacuum dominated universes). The energy of entanglement is depicted in Fig. \ref{fig:Emodos} for different values of the parameter  $w$.

It provides us with another mechanism by which the vacuum energy of the universe can be high in an earlier stage of the universe and it becomes  smaller, however, in a more evolved stage of the universe provided that the cosmic scalar field starts with a small value of the mode $k$ and evolves to higher modes as the universe is expanding.

\section{Conclusions and further comments}

It has been shown that the quantum state of a multiverse made up of homogeneous and isotropic space-times with a massless scalar field is given by a squeezed state, and that the quantum state of the phantom multiverse turns out to be an entangled state between the modes that correspond to the expanding and contracting branches of each universe, before and after the big rip singularity, respectively. A pair of entangled universes can also come up from a double instanton whose creation is allowed by the presence of quantum corrections in the Wheeler-DeWitt equation. Therefore, quantum states with no classical analog have generally to be considered in the context of the quantum multiverse.

Statistical boundary conditions have to be impose to determine the quantum state of the multiverse. The boundary condition of the multiverse that the number of universes of the multiverse does not depend on the value of the scale factor of a particular single universe fixes the representation to be chosen. This is given by the Lewis states that can be interpreted, in the context of the multiverse, as the states that represent entangled pairs of universes.

If the existence of squeezed states in the multiverse would imply a violation of Bell's inequalities, then, because there is no common space-time to the universes in the quantum multiverse, the non-locality features of the squeezed states would rather be related to the interdependence of the entangled quantum states that represent different universes or regions of the universe, being these however classically and thus causally disconnected.

It has been studied the thermodynamical properties of a closed system like the multiverse. All the thermodynamical magnitudes of a closed system are internal properties of the system and, with the given definitions, the first and second principles of thermodynamics are satisfied for any value of the scale factor. The entropy of the multiverse can decrease, satisfying however the second principle of thermodynamics  because the process is not adiabatic, the change of entropy precisely corresponds to the change of heat (over the temperature) and, thus, the entropy production is zero. 

Unlike the values of the quantum informational analogs of work and heat, the values of the quantum thermodynamical energy and entropy do not depend on the representation which is chosen to describe the state of the multiverse, provided that different representations are related to each other by unitary transformations. Therefore, if the universe starts in a pure state, such a state remains being a pure state along the unitary evolution of the universes in the multiverse.

It has also been considered a pair of universes whose quantum mechanical states are entangled. The composite state of the pair is given by a pure state. However, the state of each single universe turns out to be given by a thermal state with a temperature that depends on the scale factor. Both universes of the entangled pair stay therefore in thermal equilibrium along the correlated evolutions of their scale factors. Cosmic entanglement provides us thus with a mechanism by which the thermodynamical arrow of time in the multiverse, given by the change of the total quantum entropy, would be zero for a multiverse described in terms of pure states of entangled pairs of universes, a conclusion which could be related  to that already pointed out in Ref. \cite{Mersini2008a} [see, also, Ref. (\cite{Mersini2012})]. Each single universe of the multiverse, however, would still have an arrow of time given by the change of the entropy of entanglement with its partner universe. This arrow of time corresponds however to a decrease of the entanglement entropy rather than an increase. Nevertheless, the second principle of thermodynamics is satisfied because the change of the entropy of entanglement precisely corresponds to the change of the energy of entanglement, which can be identified with the heat of entanglement of each single universe.

The evolution of the temperature and the energy of entanglement depends on the kind of universes that are considered. For baby universes, the energy of entanglement grows with the expansion of the parent universe. It can be interpreted as an effective creation of a high number of vacuum fluctuations in the space-time of the parent universe. For parent universes, the temperature and the energy of entanglement decrease along the expansion of the universe. Thus, the energy density of each single universe can be high in an initial stage, what is expected for an inflationary period, and it can however present a smaller value in a more evolved epoch, like the current one.

The energy of entanglement for the positive and negative modes of a massless scalar field, which correspond to the expanding and contracting branches of the universe, respectively, obeys a similar behavior for the vacuum energy when the scalar field starts with a small value of the mode and evolves to higher values with the expansion of the universe.

In this paper, it has also been pointed out that the quantum-mechanical fundamental concepts of complementarity and non-locality have to be revised within the context of the quantum multiverse. Thus, multiversal non-locality has to be extended in such a way that it now expresses interdependence of different regions of the whole manifold that represents the multiverse. These regions can classically and causally be disconnected to each other although their composite state can still present quantum correlations. Thus, the classical concept of causality ought to be revised. The concept of complementarity in the multiverse implies the consideration of interference processes among different universes or branches of the universe. These processes might have observable effects in each single universe, so that it underlies the question of whether the multiverse studied in this paper can be tested, i.e. whether it is a falseable scientific proposal, actually \footnote{In words of G. Ellis \cite{Carr2007}, \emph{the issue of testability underlies the question of whether multiverse proposals are really scientific}.}.

In general, on regarding the testability of a multiverse proposal, it should be firstly said that a multiverse can actually be considered provided that it allows searching for the effects that other universes might imprint in the properties of our own universe. Furthermore, different ways of potentially observing the effects of the multiverse in our universe have been proposed. In Refs. \cite{Mersini2007, Mersini2008}, it has been proposed that giant voids in the sky could be the result of inter-universal interactions  and, in Ref. \cite{PFGD2011}, the light pattern of the gravitational lensing produced by wormholes, ringholes and Klein-bottle holes that connect our universe with others would be distinguishable from that made by similar tunnels connecting different regions of our universe, providing us thus with a mechanism for testing the multiverse.

In the model presented in this paper, it has been shown that the inter-universal entanglement can modify the dynamical and thermodynamical properties of single universes. Thus, some additional forms of testability of the quantum multiverse can be envisaged. Firstly, the temperature of entanglement might be matched with the temperature of our universe provided that there exists a relation between the thermodynamics of entanglement and the thermodynamics of the universe. Secondly, assuming that the energy of inter-universal entanglement were the major contribution to the vacuum energy of a single universe, then, the evolution rate of the scale factor would have a correlation with the amount of inter-universal entanglement and, particularly, with the rate of change of its energy of entanglement. By using the observational data, the entanglement rate could be fixed. Furthermore, different boundary conditions for the state of the multiverse imply different entanglement rates between the states of single universes. Therefore, it might well be that the evolution rate of the scale factor of our universe would provide us with a criteria for selecting the appropriate boundary condition of the whole multiverse, making testable not only the multiverse proposal but also the cosmic boundary conditions to be chosen.

Thus, the question of testability of the multiverse, far from being the problem of the multiverse, seems to be the keystone for considering new approaches to traditional questions in quantum cosmology, like the boundary conditions, the arrow of time or the anthropic principles, among others. It challenges us to adopt new and open-minded points of view about major physical and philosophical preconceptions.

\bibliographystyle{apsrev}

\bibliography{bibliography}

\begin{thebibliography}{65}
\expandafter\ifx\csname natexlab\endcsname\relax\def\natexlab#1{#1}\fi
\expandafter\ifx\csname bibnamefont\endcsname\relax
  \def\bibnamefont#1{#1}\fi
\expandafter\ifx\csname bibfnamefont\endcsname\relax
  \def\bibfnamefont#1{#1}\fi
\expandafter\ifx\csname citenamefont\endcsname\relax
  \def\citenamefont#1{#1}\fi
\expandafter\ifx\csname url\endcsname\relax
  \def\url#1{\texttt{#1}}\fi
\expandafter\ifx\csname urlprefix\endcsname\relax\def\urlprefix{URL }\fi
\providecommand{\bibinfo}[2]{#2}
\providecommand{\eprint}[2][]{\url{#2}}

\bibitem[{\citenamefont{Reid and Walls}(1986)}]{Reid1986}
\bibinfo{author}{\bibfnamefont{M.~D.} \bibnamefont{Reid}} \bibnamefont{and}
  \bibinfo{author}{\bibfnamefont{D.~F.} \bibnamefont{Walls}},
  \bibinfo{journal}{Phys. Rev. A} \textbf{\bibinfo{volume}{34}},
  \bibinfo{pages}{1260} (\bibinfo{year}{1986}).

\bibitem[{\citenamefont{Mukohyama et~al.}(1997)\citenamefont{Mukohyama, Seriu,
  and Kodama}}]{Mukohyama1997}
\bibinfo{author}{\bibfnamefont{S.}~\bibnamefont{Mukohyama}},
  \bibinfo{author}{\bibfnamefont{M.}~\bibnamefont{Seriu}}, \bibnamefont{and}
  \bibinfo{author}{\bibfnamefont{H.}~\bibnamefont{Kodama}},
  \bibinfo{journal}{Phys. Rev. D} \textbf{\bibinfo{volume}{55}},
  \bibinfo{pages}{7666} (\bibinfo{year}{1997}).

\bibitem[{\citenamefont{Mukohyama}(1998)}]{Mukohyama1998}
\bibinfo{author}{\bibfnamefont{S.}~\bibnamefont{Mukohyama}},
  \bibinfo{journal}{Phys. Rev. D} \textbf{\bibinfo{volume}{58}},
  \bibinfo{pages}{104023} (\bibinfo{year}{1998}).

\bibitem[{\citenamefont{Lee et~al.}(2007)}]{Lee2007}
\bibinfo{author}{\bibfnamefont{J.-W.} \bibnamefont{Lee}} \bibnamefont{et~al.},
  \bibinfo{journal}{JCAP} \textbf{\bibinfo{volume}{0708}}, \bibinfo{pages}{005}
  (\bibinfo{year}{2007}), \eprint{hep-th/0701199}.

\bibitem[{\citenamefont{M{\"u}ller and Lousto}(1995)}]{Muller1995}
\bibinfo{author}{\bibfnamefont{R.}~\bibnamefont{M{\"u}ller}} \bibnamefont{and}
  \bibinfo{author}{\bibfnamefont{C.~O.} \bibnamefont{Lousto}},
  \bibinfo{journal}{Phys. Rev. D} \textbf{\bibinfo{volume}{52}},
  \bibinfo{pages}{4512} (\bibinfo{year}{1995}).

\bibitem[{\citenamefont{Mersini-Houghton}(2008{\natexlab{a}})}]{Mersini2008a}
\bibinfo{author}{\bibfnamefont{L.}~\bibnamefont{Mersini-Houghton}}
  (\bibinfo{year}{2008}{\natexlab{a}}), \eprint{arXiv:0809.3623v1}.

\bibitem[{\citenamefont{Mersini-Houghton}(2008{\natexlab{b}})}]{Mersini2008b}
\bibinfo{author}{\bibfnamefont{L.}~\bibnamefont{Mersini-Houghton}}
  (\bibinfo{year}{2008}{\natexlab{b}}), \eprint{arXiv:0804.4280v1}.

\bibitem[{\citenamefont{Strominger}(1990)}]{Strominger1990}
\bibinfo{author}{\bibfnamefont{A.}~\bibnamefont{Strominger}}, in
  \emph{\bibinfo{booktitle}{Quantum Cosmology and Baby Universes}}, edited by
  \bibinfo{editor}{\bibfnamefont{S.}~\bibnamefont{Coleman}},
  \bibinfo{editor}{\bibfnamefont{J.~B.} \bibnamefont{Hartle}},
  \bibinfo{editor}{\bibfnamefont{T.}~\bibnamefont{Piran}}, \bibnamefont{and}
  \bibinfo{editor}{\bibfnamefont{S.}~\bibnamefont{Weinberg}}
  (\bibinfo{publisher}{World Scientific, London, UK}, \bibinfo{year}{1990}),
  vol.~\bibinfo{volume}{7}.

\bibitem[{\citenamefont{Robles-P{\'e}rez and
  Gonz{\'a}lez-D{\'\i}az}(2010)}]{RP2010}
\bibinfo{author}{\bibfnamefont{S.}~\bibnamefont{Robles-P{\'e}rez}}
  \bibnamefont{and} \bibinfo{author}{\bibfnamefont{P.~F.}
  \bibnamefont{Gonz{\'a}lez-D{\'\i}az}}, \bibinfo{journal}{Phys. Rev. D}
  \textbf{\bibinfo{volume}{81}}, \bibinfo{pages}{083529}
  (\bibinfo{year}{2010}), \eprint{arXiv:1005.2147v1}.

\bibitem[{\citenamefont{Vilenkin}(1986)}]{Vilenkin1986}
\bibinfo{author}{\bibfnamefont{A.}~\bibnamefont{Vilenkin}},
  \bibinfo{journal}{Phys. Rev. D} \textbf{\bibinfo{volume}{33}},
  \bibinfo{pages}{3560} (\bibinfo{year}{1986}).

\bibitem[{\citenamefont{Isham}(1992)}]{Isham1992}
\bibinfo{author}{\bibfnamefont{C.~J.} \bibnamefont{Isham}}
  (\bibinfo{year}{1992}), \eprint{arXiv:gr-qc/9210011v1}.

\bibitem[{\citenamefont{Hawking et~al.}(1992)\citenamefont{Hawking, Laflamme,
  and Lyons}}]{Hawking1992}
\bibinfo{author}{\bibfnamefont{S.~W.} \bibnamefont{Hawking}},
  \bibinfo{author}{\bibfnamefont{R.}~\bibnamefont{Laflamme}}, \bibnamefont{and}
  \bibinfo{author}{\bibfnamefont{G.~W.} \bibnamefont{Lyons}},
  \bibinfo{journal}{Phys. Rev. D} pp. \bibinfo{pages}{1546--1550}
  (\bibinfo{year}{1992}).

\bibitem[{\citenamefont{Hartle}(1995)}]{Hartle1993}
\bibinfo{author}{\bibfnamefont{J.~B.} \bibnamefont{Hartle}}, in
  \emph{\bibinfo{booktitle}{Gravitation and Quantization: Proceedings of the
  1992 Les Houches Summer School}}, edited by
  \bibinfo{editor}{\bibfnamefont{B.}~\bibnamefont{Julia}} \bibnamefont{and}
  \bibinfo{editor}{\bibfnamefont{J.}~\bibnamefont{Zinn-Justin}}
  (\bibinfo{publisher}{North Holland, Amsterdam}, \bibinfo{year}{1995}),
  \eprint{gr-qc/9304006}.

\bibitem[{\citenamefont{Kiefer and Zeh}(1995)}]{Kiefer1995}
\bibinfo{author}{\bibfnamefont{C.}~\bibnamefont{Kiefer}} \bibnamefont{and}
  \bibinfo{author}{\bibfnamefont{H.~D.} \bibnamefont{Zeh}},
  \bibinfo{journal}{Phys. Rev. D} \textbf{\bibinfo{volume}{51}},
  \bibinfo{pages}{4145} (\bibinfo{year}{1995}).

\bibitem[{\citenamefont{Kiefer}(2005)}]{Kiefer2005}
\bibinfo{author}{\bibfnamefont{C.}~\bibnamefont{Kiefer}},
  \bibinfo{journal}{Braz. J. Phys.} \textbf{\bibinfo{volume}{35}},
  \bibinfo{pages}{296} (\bibinfo{year}{2005}).

\bibitem[{\citenamefont{Kiefer}(2007)}]{Kiefer2007}
\bibinfo{author}{\bibfnamefont{C.}~\bibnamefont{Kiefer}},
  \emph{\bibinfo{title}{Quantum gravity}} (\bibinfo{publisher}{Oxford
  University Press, Oxford, UK}, \bibinfo{year}{2007}).

\bibitem[{\citenamefont{De~Witt}(1967)}]{DeWitt1967}
\bibinfo{author}{\bibfnamefont{B.~S.} \bibnamefont{De~Witt}},
  \bibinfo{journal}{Phys. Rev.} \textbf{\bibinfo{volume}{160}},
  \bibinfo{pages}{1113} (\bibinfo{year}{1967}).

\bibitem[{\citenamefont{Wiltshire}(2003)}]{Wiltshire2003}
\bibinfo{author}{\bibfnamefont{D.~L.} \bibnamefont{Wiltshire}}
  (\bibinfo{year}{2003}), \eprint{gr-qc/0101003}.

\bibitem[{\citenamefont{Vilenkin}(1982)}]{Vilenkin1982}
\bibinfo{author}{\bibfnamefont{A.}~\bibnamefont{Vilenkin}},
  \bibinfo{journal}{Phys. Lett. B} \textbf{\bibinfo{volume}{117}},
  \bibinfo{pages}{25} (\bibinfo{year}{1982}).

\bibitem[{\citenamefont{Hawking}(1984)}]{Hawking1983}
\bibinfo{author}{\bibfnamefont{S.~W.} \bibnamefont{Hawking}}, in
  \emph{\bibinfo{booktitle}{Relativity, groups and topology II, Les Houches,
  Session XL, 1983}}, edited by \bibinfo{editor}{\bibfnamefont{B.~S.}
  \bibnamefont{De~Witt}} \bibnamefont{and}
  \bibinfo{editor}{\bibfnamefont{R.}~\bibnamefont{Stora}}
  (\bibinfo{publisher}{Elsevier Science Publishers B. V.},
  \bibinfo{year}{1984}).

\bibitem[{\citenamefont{Barvinsky and Kamenshchik}(2006)}]{Barvinsky2006}
\bibinfo{author}{\bibfnamefont{A.~O.} \bibnamefont{Barvinsky}}
  \bibnamefont{and} \bibinfo{author}{\bibfnamefont{A.~Y.}
  \bibnamefont{Kamenshchik}}, \bibinfo{journal}{JCAP}
  \textbf{\bibinfo{volume}{0609}}, \bibinfo{pages}{014} (\bibinfo{year}{2006}),
  \eprint{hep-th/0605132}.

\bibitem[{\citenamefont{Barvinsky and Kamenshchik}(2007)}]{Barvinsky2007a}
\bibinfo{author}{\bibfnamefont{A.~O.} \bibnamefont{Barvinsky}}
  \bibnamefont{and} \bibinfo{author}{\bibfnamefont{A.~Y.}
  \bibnamefont{Kamenshchik}}, \bibinfo{journal}{J. Phys. A}
  \textbf{\bibinfo{volume}{40}}, \bibinfo{pages}{7043} (\bibinfo{year}{2007}),
  \eprint{hep-th/0701201}.

\bibitem[{\citenamefont{Barvinsky}(2007)}]{Barvinsky2007b}
\bibinfo{author}{\bibfnamefont{A.~O.} \bibnamefont{Barvinsky}},
  \bibinfo{journal}{Phys. Rev. Lett.} \textbf{\bibinfo{volume}{99}},
  \bibinfo{pages}{071301} (\bibinfo{year}{2007}), \eprint{0704.0083}.

\bibitem[{\citenamefont{Halliwell}(1987)}]{Halliwell1987}
\bibinfo{author}{\bibfnamefont{J.~J.} \bibnamefont{Halliwell}},
  \bibinfo{journal}{Phys. Rev. D} \textbf{\bibinfo{volume}{36}},
  \bibinfo{pages}{3626} (\bibinfo{year}{1987}).

\bibitem[{\citenamefont{Hawking}(1982)}]{Hawking1982}
\bibinfo{author}{\bibfnamefont{S.~W.} \bibnamefont{Hawking}},
  \bibinfo{journal}{Astrophysical Cosmology, 563-72. Vatican City: Pontificia
  Academiae Scientarium}  (\bibinfo{year}{1982}).

\bibitem[{\citenamefont{Halliwell}(1990)}]{Halliwell1990}
\bibinfo{author}{\bibfnamefont{J.~J.} \bibnamefont{Halliwell}}, in
  \emph{\bibinfo{booktitle}{Quantum Cosmology and Baby Universes}}, edited by
  \bibinfo{editor}{\bibfnamefont{S.}~\bibnamefont{Coleman}},
  \bibinfo{editor}{\bibfnamefont{J.~B.} \bibnamefont{Hartle}},
  \bibinfo{editor}{\bibfnamefont{T.}~\bibnamefont{Piran}}, \bibnamefont{and}
  \bibinfo{editor}{\bibfnamefont{S.}~\bibnamefont{Weinberg}}
  (\bibinfo{publisher}{World Scientific, London, UK}, \bibinfo{year}{1990}),
  vol.~\bibinfo{volume}{7}.

\bibitem[{\citenamefont{Hawking}(1985)}]{Hawking1985}
\bibinfo{author}{\bibfnamefont{S.~W.} \bibnamefont{Hawking}},
  \bibinfo{journal}{Phys. Rev. D} \textbf{\bibinfo{volume}{32}},
  \bibinfo{pages}{2489} (\bibinfo{year}{1985}).

\bibitem[{\citenamefont{Halliwell}(1989)}]{Halliwell1989}
\bibinfo{author}{\bibfnamefont{J.~J.} \bibnamefont{Halliwell}},
  \bibinfo{journal}{Phys. Rev. D} \textbf{\bibinfo{volume}{39}},
  \bibinfo{pages}{2912} (\bibinfo{year}{1989}).

\bibitem[{\citenamefont{Kiefer}(1992)}]{Kiefer1992}
\bibinfo{author}{\bibfnamefont{C.}~\bibnamefont{Kiefer}},
  \bibinfo{journal}{Phys. Rev. D} \textbf{\bibinfo{volume}{46}},
  \bibinfo{pages}{1658} (\bibinfo{year}{1992}).

\bibitem[{\citenamefont{Robles-P{\'e}rez
  et~al.}(2011)\citenamefont{Robles-P{\'e}rez, Alonso-Serrano, and
  Gonz{\'a}lez-D{\'\i}az}}]{RP2011}
\bibinfo{author}{\bibfnamefont{S.}~\bibnamefont{Robles-P{\'e}rez}},
  \bibinfo{author}{\bibfnamefont{A.}~\bibnamefont{Alonso-Serrano}},
  \bibnamefont{and} \bibinfo{author}{\bibfnamefont{P.~F.}
  \bibnamefont{Gonz{\'a}lez-D{\'\i}az}}, \bibinfo{journal}{Phys. Rev. D}
  \textbf{\bibinfo{volume}{85}}, \bibinfo{pages}{063611}
  (\bibinfo{year}{2011}), \eprint{arXiv:1111.3178}.

\bibitem[{\citenamefont{Abramovitz and Stegun}(1972)}]{Abramovitz1972}
\bibinfo{editor}{\bibfnamefont{M.}~\bibnamefont{Abramovitz}} \bibnamefont{and}
  \bibinfo{editor}{\bibfnamefont{I.~A.} \bibnamefont{Stegun}}, eds.,
  \emph{\bibinfo{title}{Handbook of Mathematical Functions}}
  (\bibinfo{publisher}{NBS}, \bibinfo{year}{1972}).

\bibitem[{\citenamefont{Birrell and Davies}(1982)}]{Birrell1982}
\bibinfo{author}{\bibfnamefont{N.~D.} \bibnamefont{Birrell}} \bibnamefont{and}
  \bibinfo{author}{\bibfnamefont{P.~C.~W.} \bibnamefont{Davies}},
  \emph{\bibinfo{title}{Quantum fields in curved space}}
  (\bibinfo{publisher}{Cambridge University Press, Cambridge, UK},
  \bibinfo{year}{1982}).

\bibitem[{\citenamefont{Gonz{\'a}lez-D{\'\i}az and
  Robles-P{\'e}rez}(2008)}]{PFGD2007}
\bibinfo{author}{\bibfnamefont{P.~F.} \bibnamefont{Gonz{\'a}lez-D{\'\i}az}}
  \bibnamefont{and}
  \bibinfo{author}{\bibfnamefont{S.}~\bibnamefont{Robles-P{\'e}rez}},
  \bibinfo{journal}{Int. J. Mod. Phys. D} \textbf{\bibinfo{volume}{17}},
  \bibinfo{pages}{1213} (\bibinfo{year}{2008}), \eprint{0709.4038}.

\bibitem[{\citenamefont{Caldwell}(2002)}]{Caldwell2002}
\bibinfo{author}{\bibfnamefont{R.~R.} \bibnamefont{Caldwell}},
  \bibinfo{journal}{Phys. Lett. B} \textbf{\bibinfo{volume}{545}},
  \bibinfo{pages}{23} (\bibinfo{year}{2002}), \eprint{astro-ph/9908168}.

\bibitem[{\citenamefont{Caldwell and Kamionkowski}(2009)}]{Caldwell2009}
\bibinfo{author}{\bibfnamefont{R.~R.} \bibnamefont{Caldwell}} \bibnamefont{and}
  \bibinfo{author}{\bibfnamefont{M.}~\bibnamefont{Kamionkowski}},
  \bibinfo{journal}{Ann. Rev. Nucl. Part. Sci.} \textbf{\bibinfo{volume}{59}},
  \bibinfo{pages}{397} (\bibinfo{year}{2009}).

\bibitem[{\citenamefont{Nojiri and Odintsov}(2004)}]{Nojiri2004}
\bibinfo{author}{\bibfnamefont{S.}~\bibnamefont{Nojiri}} \bibnamefont{and}
  \bibinfo{author}{\bibfnamefont{S.~D.} \bibnamefont{Odintsov}},
  \bibinfo{journal}{Phys. Rev. D} \textbf{\bibinfo{volume}{70}},
  \bibinfo{pages}{103522} (\bibinfo{year}{2004}).

\bibitem[{\citenamefont{Gonz{\'a}lez-D{\'\i}az}(2003)}]{PFGD2003}
\bibinfo{author}{\bibfnamefont{P.~F.} \bibnamefont{Gonz{\'a}lez-D{\'\i}az}},
  \bibinfo{journal}{Phys. Rev. D} \textbf{\bibinfo{volume}{68}},
  \bibinfo{pages}{084016} (\bibinfo{year}{2003}).

\bibitem[{\citenamefont{Lobo}(2005)}]{Lobo2005}
\bibinfo{author}{\bibfnamefont{F.~S.~N.} \bibnamefont{Lobo}},
  \bibinfo{journal}{Phys. Rev. D} \textbf{\bibinfo{volume}{71}},
  \bibinfo{pages}{084011} (\bibinfo{year}{2005}).

\bibitem[{\citenamefont{Lewis and Riesenfeld}(1969)}]{Lewis1969}
\bibinfo{author}{\bibfnamefont{H.~R.} \bibnamefont{Lewis}} \bibnamefont{and}
  \bibinfo{author}{\bibfnamefont{W.~B.} \bibnamefont{Riesenfeld}},
  \bibinfo{journal}{J. Math. Phys.} \textbf{\bibinfo{volume}{10}},
  \bibinfo{pages}{1458} (\bibinfo{year}{1969}).

\bibitem[{\citenamefont{Pedrosa}(1987)}]{Pedrosa1987}
\bibinfo{author}{\bibfnamefont{I.~A.} \bibnamefont{Pedrosa}},
  \bibinfo{journal}{Phys. Rev. D} \textbf{\bibinfo{volume}{36}},
  \bibinfo{pages}{1279} (\bibinfo{year}{1987}).

\bibitem[{\citenamefont{Dantas et~al.}(1992)\citenamefont{Dantas, Pedrosa, and
  Baseia}}]{Dantas1992}
\bibinfo{author}{\bibfnamefont{C.~M.~A.} \bibnamefont{Dantas}},
  \bibinfo{author}{\bibfnamefont{I.~A.} \bibnamefont{Pedrosa}},
  \bibnamefont{and} \bibinfo{author}{\bibfnamefont{B.}~\bibnamefont{Baseia}},
  \bibinfo{journal}{Phys. Rev. A} \textbf{\bibinfo{volume}{45}},
  \bibinfo{pages}{1320} (\bibinfo{year}{1992}).

\bibitem[{\citenamefont{Sheng et~al.}(1995)}]{Sheng1995}
\bibinfo{author}{\bibfnamefont{D.}~\bibnamefont{Sheng}} \bibnamefont{et~al.},
  \bibinfo{journal}{Int. J. Theor. Phys.} \textbf{\bibinfo{volume}{34}},
  \bibinfo{pages}{355} (\bibinfo{year}{1995}).

\bibitem[{\citenamefont{Song}(2000)}]{Song2000}
\bibinfo{author}{\bibfnamefont{D.-Y.} \bibnamefont{Song}},
  \bibinfo{journal}{Phys. Rev. A} \textbf{\bibinfo{volume}{62}},
  \bibinfo{pages}{014103} (\bibinfo{year}{2000}).

\bibitem[{\citenamefont{Park}(2004)}]{Park2004}
\bibinfo{author}{\bibfnamefont{T.~J.} \bibnamefont{Park}},
  \bibinfo{journal}{Bull. Korean Chem. Soc.} \textbf{\bibinfo{volume}{25}}
  (\bibinfo{year}{2004}).

\bibitem[{\citenamefont{Vergel and Villase{\~n}or}(2009)}]{Vergel2009}
\bibinfo{author}{\bibfnamefont{D.~G.} \bibnamefont{Vergel}} \bibnamefont{and}
  \bibinfo{author}{\bibfnamefont{J.~S.} \bibnamefont{Villase{\~n}or}},
  \bibinfo{journal}{Ann. Phys.} \textbf{\bibinfo{volume}{324}},
  \bibinfo{pages}{1360} (\bibinfo{year}{2009}).

\bibitem[{\citenamefont{Yuen}(1975)}]{Yuen1975}
\bibinfo{author}{\bibfnamefont{H.~P.} \bibnamefont{Yuen}},
  \bibinfo{journal}{Phys. Lett. A} \textbf{\bibinfo{volume}{51}},
  \bibinfo{pages}{1} (\bibinfo{year}{1975}).

\bibitem[{\citenamefont{Yuen}(1976)}]{Yuen1976}
\bibinfo{author}{\bibfnamefont{H.~P.} \bibnamefont{Yuen}},
  \bibinfo{journal}{Phys. Rev. A} \textbf{\bibinfo{volume}{13}},
  \bibinfo{pages}{2226} (\bibinfo{year}{1976}).

\bibitem[{\citenamefont{Robles-P{\'e}rez
  et~al.}(2010)\citenamefont{Robles-P{\'e}rez, Hassouni, and
  Gonz{\'a}lez-D{\'\i}az}}]{RP2009}
\bibinfo{author}{\bibfnamefont{S.}~\bibnamefont{Robles-P{\'e}rez}},
  \bibinfo{author}{\bibfnamefont{Y.}~\bibnamefont{Hassouni}}, \bibnamefont{and}
  \bibinfo{author}{\bibfnamefont{P.~F.} \bibnamefont{Gonz{\'a}lez-D{\'\i}az}},
  \bibinfo{journal}{Phys. Lett. B} \textbf{\bibinfo{volume}{683}},
  \bibinfo{pages}{1} (\bibinfo{year}{2010}), \eprint{0909.3063}.

\bibitem[{\citenamefont{Wheeler}(1957)}]{Wheeler1957}
\bibinfo{author}{\bibfnamefont{J.~A.} \bibnamefont{Wheeler}},
  \bibinfo{journal}{Ann. Phys.} \textbf{\bibinfo{volume}{2}},
  \bibinfo{pages}{604} (\bibinfo{year}{1957}).

\bibitem[{\citenamefont{Grishchuk and Sidorov}(1990)}]{Grishchuk1990}
\bibinfo{author}{\bibfnamefont{L.~P.} \bibnamefont{Grishchuk}}
  \bibnamefont{and} \bibinfo{author}{\bibfnamefont{Y.~V.}
  \bibnamefont{Sidorov}}, \bibinfo{journal}{Phys. Rev. D}
  \textbf{\bibinfo{volume}{42}}, \bibinfo{pages}{3413} (\bibinfo{year}{1990}).

\bibitem[{\citenamefont{Scully and Zubairy}(1997)}]{Scully1997}
\bibinfo{author}{\bibfnamefont{M.~O.} \bibnamefont{Scully}} \bibnamefont{and}
  \bibinfo{author}{\bibfnamefont{M.~S.} \bibnamefont{Zubairy}},
  \emph{\bibinfo{title}{Quantum optics}} (\bibinfo{publisher}{Cambridge
  University Press, Cambridge, UK}, \bibinfo{year}{1997}).

\bibitem[{\citenamefont{Einstein et~al.}(1935)}]{Einstein1935}
\bibinfo{author}{\bibfnamefont{A.}~\bibnamefont{Einstein}}
  \bibnamefont{et~al.}, \bibinfo{journal}{Phys. Rev.}
  \textbf{\bibinfo{volume}{47}}, \bibinfo{pages}{777} (\bibinfo{year}{1935}).

\bibitem[{\citenamefont{Bell}(1987)}]{Bell1987}
\bibinfo{author}{\bibfnamefont{J.~S.} \bibnamefont{Bell}},
  \emph{\bibinfo{title}{Speakable and unspeakable in quantum mechanics}}
  (\bibinfo{publisher}{Cambridge University Press, Cambridge, UK},
  \bibinfo{year}{1987}).

\bibitem[{\citenamefont{Alicki et~al.}(2004)}]{Alicki2004}
\bibinfo{author}{\bibfnamefont{R.}~\bibnamefont{Alicki}} \bibnamefont{et~al.},
  \bibinfo{journal}{Open Syst. Inf. Dyn.} \textbf{\bibinfo{volume}{11}},
  \bibinfo{pages}{205} (\bibinfo{year}{2004}),
  \eprint{arXiv:quant-ph/0402012v2}.

\bibitem[{\citenamefont{Gemmer et~al.}(2009)}]{Gemmer2009}
\bibinfo{author}{\bibfnamefont{J.}~\bibnamefont{Gemmer}} \bibnamefont{et~al.},
  \emph{\bibinfo{title}{Quantum thermodynamics}}
  (\bibinfo{publisher}{Springer-Verlag, Berlin, Germany},
  \bibinfo{year}{2009}).

\bibitem[{\citenamefont{Joos et~al.}(2003)}]{Joos2003}
\bibinfo{author}{\bibfnamefont{E.}~\bibnamefont{Joos}} \bibnamefont{et~al.},
  \emph{\bibinfo{title}{Decoherence and the Appearance of a Classical World in
  Quantum Theory}} (\bibinfo{publisher}{Springer-Verlag, Berlin, Germany},
  \bibinfo{year}{2003}).

\bibitem[{\citenamefont{Vedral}(2006)}]{Vedral2006}
\bibinfo{author}{\bibfnamefont{V.}~\bibnamefont{Vedral}},
  \emph{\bibinfo{title}{Introduction to quantum information science}}
  (\bibinfo{publisher}{Oxford University Press, Oxford, UK},
  \bibinfo{year}{2006}).

\bibitem[{\citenamefont{Wodkiewicz and Eberly}(1985)}]{Wodkiewicz1985}
\bibinfo{author}{\bibfnamefont{K.}~\bibnamefont{Wodkiewicz}} \bibnamefont{and}
  \bibinfo{author}{\bibfnamefont{J.~H.} \bibnamefont{Eberly}},
  \bibinfo{journal}{J. Opt. Soc. Am. B} \textbf{\bibinfo{volume}{2}},
  \bibinfo{pages}{458} (\bibinfo{year}{1985}).

\bibitem[{\citenamefont{Buzek}(1989)}]{Buzek1989}
\bibinfo{author}{\bibfnamefont{V.}~\bibnamefont{Buzek}},
  \bibinfo{journal}{Phys. Lett. A} \textbf{\bibinfo{volume}{139}},
  \bibinfo{pages}{231} (\bibinfo{year}{1989}).

\bibitem[{\citenamefont{Mukhanov and Winitzki}(2007)}]{Mukhanov2007}
\bibinfo{author}{\bibfnamefont{V.~F.} \bibnamefont{Mukhanov}} \bibnamefont{and}
  \bibinfo{author}{\bibfnamefont{S.}~\bibnamefont{Winitzki}},
  \emph{\bibinfo{title}{Quantum Effects in Gravity}}
  (\bibinfo{publisher}{Cambridge University Press, Cambridge, UK},
  \bibinfo{year}{2007}).

\bibitem[{\citenamefont{Mersini-Houghton}(2012)}]{Mersini2012}
\bibinfo{author}{\bibfnamefont{L.}~\bibnamefont{Mersini-Houghton}},
  \emph{\bibinfo{title}{The arrows of time: a debate in cosmology}}
  (\bibinfo{publisher}{Springer, Berlin, Germany}, \bibinfo{year}{2012}).

\bibitem[{\citenamefont{Mersini-Houghton}(2007)}]{Mersini2007}
\bibinfo{author}{\bibfnamefont{L.}~\bibnamefont{Mersini-Houghton}},
  \bibinfo{journal}{New Scientist} pp. \bibinfo{pages}{11--24}
  (\bibinfo{year}{2007}).

\bibitem[{\citenamefont{Holman et~al.}(2008)\citenamefont{Holman,
  Mersini-Houghton, and Takahashi}}]{Mersini2008}
\bibinfo{author}{\bibfnamefont{R.}~\bibnamefont{Holman}},
  \bibinfo{author}{\bibfnamefont{L.}~\bibnamefont{Mersini-Houghton}},
  \bibnamefont{and}
  \bibinfo{author}{\bibfnamefont{T.}~\bibnamefont{Takahashi}},
  \bibinfo{journal}{Phys. Rev. D} \textbf{\bibinfo{volume}{77}},
  \bibinfo{pages}{063510,063511} (\bibinfo{year}{2008}),
  \eprint{[arXiv:hep-th/0611223v1], [arXiv:hep-th/0612142v1]}.

\bibitem[{\citenamefont{Gonz{\'a}lez-D{\'\i}az and
  Alonso-Serrano}(2011)}]{PFGD2011}
\bibinfo{author}{\bibfnamefont{P.~F.} \bibnamefont{Gonz{\'a}lez-D{\'\i}az}}
  \bibnamefont{and}
  \bibinfo{author}{\bibfnamefont{A.}~\bibnamefont{Alonso-Serrano}},
  \bibinfo{journal}{Phys. Rev. D} \textbf{\bibinfo{volume}{84}},
  \bibinfo{pages}{023008} (\bibinfo{year}{2011}).

\bibitem[{\citenamefont{Carr}(2007)}]{Carr2007}
\bibinfo{editor}{\bibfnamefont{B.}~\bibnamefont{Carr}}, ed.,
  \emph{\bibinfo{title}{Universe or Multiverse}} (\bibinfo{publisher}{Cambridge
  University Press, Cambridge, UK}, \bibinfo{year}{2007}).

\end{thebibliography}

\end{document}